\documentclass[a4paper,11pt]{article}
\pdfoutput=1 

\usepackage{jheppub} 

\usepackage{amssymb}
\usepackage{amsmath}
\usepackage{tikz}
\usepackage{graphicx}
\definecolor{uibred}{RGB}{167, 38, 47}

\newcommand{\xt}{\mathbf{x}}
\newcommand{\xd}{\overline{\mathbf{x}}}

\newcommand{\yt}{\mathbf{y}}
\newcommand{\yd}{\overline{\mathbf{y}}}

\newcommand{\zt}{\mathbf{z}}

\newcommand{\bt}{\mathbf{b}}
\newcommand{\rt}{\mathbf{r}}

\newcommand{\kt}{\mathbf{k}}
\newcommand{\ktt}{|\mathbf{k}|}

\newcommand{\kd}{\overline{\mathbf{k}}}

\newcommand{\nabt}{\boldsymbol{\nabla}}

\newcommand{\pt}{\mathbf{p}}

\newcommand{\pd}{\overline{\mathbf{p}}}

\newcommand{\qt}{\mathbf{q}}

\newcommand{\qd}{\overline{\mathbf{q}}}

\newcommand{\lt}{\mathbf{l}}
\newcommand{\ld}{\overline{\mathbf{l}}}
\newcommand{\nr}[1]{(\ref{#1})} 
\newcommand{\Pt}{\mathbf{P}}
\newcommand{\Qt}{\mathbf{Q}}

\newcommand{\ud}{\mathrm{d}}
\newcommand{\nc}{N_\mathrm{c}}
\newcommand{\cf}{C_\mathrm{F}}
\newcommand{\qs}{Q_\mathrm{s}}
\newcommand{\as}{\alpha_\mathrm{s}}

\newcommand{\tr}{\text{tr}}

\newcommand{\nudot}{\dot{\nu}}

\begin{document}

\title{Linearly polarized gluons and axial charge fluctuations in the Glasma}

\author[a,b]{Tuomas Lappi}
\emailAdd{tuomas.v.v.lappi@jyu.fi}
\affiliation[a]{Department of Physics, P.O. Box 35, 40014 University of Jyv\"{a}skyl\"{a}, Finland}
\affiliation[b]{Helsinki Institute of Physics, P.O. Box 64, 00014 University of Helsinki, Finland}

\author[c]{S\"{o}ren Schlichting}
\emailAdd{sslng@uw.edu}
\affiliation[c]{Department of Physics, University of Washington, Seattle, WA 98195-1560, USA}


\abstract{
We calculate of the one- and two-point correlation functions of the energy density and the divergence of the Chern-Simons current in the nonequilibrium Glasma state formed in a high-energy nuclear collision. We show that the latter depends on the difference of the total and linearly polarized gluon transverse momentum distributions. Since the divergence of the Chern-Simons current provides the source of axial charge, we infer information about the statistical properties of axial charge production at early times. We further develop a simple phenomenological model to characterize axial charge distributions in terms of distributions of the energy density.}

\date{\today}
\maketitle
\flushbottom

\section{Introduction}
Novel transport phenomena associated with the Chiral magnetic~\cite{Fukushima:2008xe,Vilenkin:1980fu,Son:2004tq} and related effects have recently caused an excitement across different fields of physics. In the high-energy QCD context, experimental measurements at RHIC and LHC  have provided intriguing hints at possible signatures of such anomalous transport phenomena  ~\cite{Abelev:2009ac,Abelev:2012pa,Adamczyk:2014mzf,Tribedy:2017hwn}. However, the interpretation of these experimental results remains inconclusive~\cite{Kharzeev:2015znc,Skokov:2016yrj} due to the presence of large background effects~\cite{Voloshin:2004vk,Wang:2009kd,Schlichting:2010qia,Pratt:2010zn,Bzdak:2010fd}.  Despite significant progress on the theory side in developing different microscopic~\cite{Stephanov:2012ki,Gao:2012ix,Son:2012zy,Mueller:2016ven,Mace:2016shq,Mueller:2017arw} and macroscopic~\cite{Newman:2005hd,Son:2009tf} descriptions of the coupled dynamics of vector and axial charges, a first principles description of the effects in high-energy heavy-ion collisions remains an outstanding challenge. Present phenomenological predictions \cite{Hirono:2014oda,Yin:2015fca,Hirono:2016lgg,Jiang:2016wve} have to rely to a varying extent on modeling assumptions. Most importantly, all phenomenological descriptions based e.g. on anomalous hydrodynamics~\cite{Newman:2005hd,Son:2009tf} require information about the early time dynamics as an initial condition for the subsequent space-time evolution. Even though significant progress has been achieved in understanding the early time dynamics of the conserved energy-momentum tensor, both from a theoretical perspective \cite{Schenke:2012wb,Kurkela:2017hgm} as well as through sophisticated model/data comparisons \cite{Gale:2012rq,Bernhard:2016tnd}, achieving a similar level of understanding of the space-time dynamics of axial charge production and anomalous transport processes during the very early pre-equilibrium stages  ($\lesssim 1$fm$/c$) remains a key challenge. 

One important difference between the dynamics of vector and axial charges, is the fact that the density of axial charge is \emph{not} conserved. This is due to the axial anomaly, which for $N_f$ flavors of (approximately) massless fermions takes the form 
\begin{eqnarray}
\label{eq:Anomaly}
\partial_{\mu}j^{\mu}_{(5)}=-\frac{g^2 N_f}{8\pi^2} \tr\Big( F_{\mu\nu}\tilde{F}^{\mu\nu} \Big) = -\frac{g^2 N_f}{16\pi^2}  F^a_{\mu\nu}\tilde{F}^{a,\mu\nu},
\end{eqnarray}
where $F_{\mu\nu}$ denotes the field strength and $\tilde{F}^{\mu\nu} = \frac{1}{2}\varepsilon^{\mu\nu\rho\sigma}F_{\rho\sigma}$ its dual. Hence understanding the dynamics of axial charges and currents in a QCD plasma inevitably requires some knowledge about the structure of non-abelian gauge fields entering on the right-hand side of Eq.~(\ref{eq:Anomaly}). Even though it is understood that in the long time and long wave-length limit, topological (sphaleron) transitions dominate the production/dissociation of axial charge~(see e.g. \cite{Moore:2010jd} and references therein), it is not clear to what extent these considerations apply to the typical time and length scales relevant during the early stages of high-energy heavy-ion collisions~\cite{Mace:2016svc}. Despite the fact that the rate of topological transitions can be significantly enhanced during the early time pre-equilibrium stage \cite{Mace:2016svc}, various kinds of short distance field strength fluctuations can also contribute significantly to axial charge production at early times. Consequently, it is of crucial importance to understand different mechanisms of axial charge production in order to estimate the magnitude and features and isolate the most relevant effects. 

One more direct way to study the strong gauge fields that dominate the initial stages of heavy ion collisions is to probe them with a dilute probe, such as in high energy deep inelastic scattering. The experimental program at a future Electron-Ion Collider~\cite{Accardi:2012qut} will be able to characterize the spacetime structure of partons inside nucleons and nuclei in a variety of ways. Out of these the linearly polarized gluon transverse momentum distribution~\cite{Mulders:2000sh,Meissner:2007rx} has recently been of particular interest to the small-$x$ community. Based on the Color-Glass-Condensate (CGC) picture it has been shown~\cite{Metz:2011wb,Dominguez:2011br,Dumitru:2015gaa} that the linearly polarized gluon distribution can be related to, and ultimately calculated from, the same Wilson line correlators that characterize unpolarized gluon distributions. We will show in this paper, that the correlation structure of the gauge fields at the earliest times after a heavy ion collision is sensitive to both the linearly polarized and unpolarized gluon distributions. It turns out that the correlations and fluctuations of axial charge are particularly sensitive to the polarized distributions. Whereas for energy density fluctuations the polarized and unpolarized contributions add up, for the axial charge they appear with a different sign. This observation opens up a fascinating new connection between correlation studies in deep inelastic scattering and local CP-violating fluctuations in hadronic collisions.

The aim of this paper is to calculate the statistical properties of axial charge production at the earliest stages of a high energy heavy ion collision. The calculation is based on the description of the early time dynamics in the Color Glass Condensate framework~\cite{Iancu:2003xm,Weigert:2005us,Gelis:2010nm,Gelis:2015gza}, which leads to the presence of  longitudinal chromo-electric and chromo-magnetic fields at very early times after the collision. We start with brief  discussion of the space-time structure of chromo-electric and chromo-magnetic fields at very early times in Sec.~\ref{sec:fluxtubes}.  We then review in Sec.~\ref{sec:correlators} the structure of the linearly polarized and unpolarized Weisz\"{a}cker-Williams (WW) gluon distributions in the CGC framework. With a Gaussian approximation for the field correlators (the ``glasma graph'' approximation) we then perform an analytic calculation of the one and two-point correlation functions of the energy density $\varepsilon(x)$ and the divergence of the Chern-Simons current $\nudot(x)$ in terms of the WW correlators. Here our calculation generalizes the closely related earlier work of~\cite{Muller:2011bb}. 
We then in Sec.~\ref{sec:momspace} relate our calculation to works studying two-gluon correlations using the glasma graph approximation. We finally discuss the implications of our results for the basic phenomenological  properties of axial charge production in the Glasma in Sec.~\ref{sec:phenomenology}, developing a simple algorithm for using our results in anomalous hydrodynamical calculations. We conclude in Sec.~\ref{sec:conc} with a summary of our results and perspectives for future studies.

\section{Glasma flux tubes and axial charge production}
\label{sec:fluxtubes}
The CGC effective theory description of a high energy nucleon or nucleus is based on a a separation of scales. Degrees of freedom carrying a large fraction of the energy of the projectile/target are described as a color charge, which acts as a source for the small-$x$ gluons. The color field of a single nucleus can be expressed analytically in terms of the color charges. When transformed to light cone gauge, these fields 
(which we denote here by $\alpha^i$ and $\beta^i$ for the two nuclei)
are  ``transverse pure gauge'' fields~\cite{McLerran:1993ka,McLerran:1994vd}
\begin{equation}
\alpha^{i}_{\xt}=\frac{i}{g} U_{\xt} \partial^{i} U_{\xt}^{\dagger}\;, \qquad \beta^{i}_{\xt}=\frac{i}{g} V_{\xt} \partial^{i} V_{\xt}^{\dagger}\;.
\end{equation}
Here $U_{\xt}$ and $V_{\xt}$ are light-like Wilson lines, which are scattering amplitudes for the eikonal interaction of a color charge passing through the color field.

Based on this picture, a high-energy heavy ion collision is realized when two such systems pass though each other. In this case the color fields of the projectile and target interact with each other, leading to the formation of a non-equilibirum ``Glasma''~\cite{Lappi:2006fp} state. By requiring that the fields be continuous over the future light cone one obtains~\cite{Kovner:1995ja,Kovner:1995ts} the gauge fields immediately after the collision at $\tau=0^{+}$ as
\begin{equation}\label{eq:potic}
A^i_{\xt} = \alpha^i_{\xt} + \beta^i_{\xt} \;,
\qquad
A^\eta = \frac{ig}{2}[\alpha^i_{\xt},\beta^i_{\xt}] \; .
\end{equation}
In terms of the field strength tensor these correspond to longitudinal color-electric and color-magnetic fields~\cite{Kovner:1995ja,Kovner:1995ts,Lappi:2006fp,Schenke:2012wb,Schenke:2012hg}
\begin{equation}\label{eq:fieldic}
E^{\eta}_{\xt}= -ig \delta^{ij} [\alpha^{i}_{\xt},\beta^{j}_{\xt}]\;, 
\qquad B^{\eta}_{\xt}= -ig\varepsilon^{ij} [\alpha^{i}_{\xt},\beta^{j}_{\xt}]\;.
\end{equation}
We use the  sign convention $D_\mu = \partial_\mu + i g A_\mu$ for the covariant derivative, and take the electric and magnetic fields in usual Minkowski coordinates to be $E^i \equiv F_{0i}$ and $B^i = - \frac{1}{2}\varepsilon^{ijk}F_{jk}$. For the components in proper-time rapidity coordinates we define $E^\eta \equiv \frac{1}{\tau}F_{\tau \eta}$, $E^i = F_{\tau i},$ $B^\eta \equiv - \frac{1}{2}\varepsilon^{ij}F_{ij}$ and $B^i = -\varepsilon^{ij}\frac{1}{\tau}F_{j\eta}$,
which at midrapidity reduce to the fields in Minkowski coordinates $E^\eta|_{\eta=0} = E^z \equiv F_{tz}$ etc.

The subsequent dynamics at very early times has been studied in great detail analytically e.g. based on small proper time expansions~\cite{Chen:2013ksa,Chen:2015wia}, as well as numerically through real-time lattice simulations~\cite{Krasnitz:1998ns,Krasnitz:1999wc,Krasnitz:2000gz,Krasnitz:2001qu,Lappi:2003bi,Gelfand:2016yho,Ipp:2017lho}. On a time scale $\tau \sim 1/\qs$ the classical Yang-Mills dynamics leads to the decoherence of the longitudinal fields building up transverse field strengths $E^{i}$ and $B^{i}$. Eventually the longitudinal expansion leads to a significant reduction of the field strength, where the semi-classical description becomes inapplicable~\cite{Berges:2013eia,Berges:2013fga,Berges:2014yta} and the system undergoes a kinetic regime before approaching local thermal equilibrium~\cite{Kurkela:2015qoa,Keegan:2016cpi}.

Even though the structure of the boost-invariant fields in Eq.~\nr{eq:fieldic} is topologically trivial~\cite{Kharzeev:2001ev}, the strong longitudinal chromo electric and chromo electric fields at early times can still contribute significantly to axial charge production. Despite the fact that the axial charge is of course carried by the fermionic degrees of freedom, an estimate of this effect can be immediately deduced from the axial anomaly relation. In this spirit, a first estimate of the fluctuations of the net axial charge density per unit rapidity 
\begin{eqnarray}
\frac{\ud N_5}{\ud\eta}\equiv \int d^2\xt~\tau~j^{\tau}_{(5)}(\xt)\;,
\end{eqnarray}
was provided in Ref.~\cite{Kharzeev:2001ev} based on explicit numerical simulations of the early time dynamics (see also \cite{Hirono:2014oda} for a parametric estimate used in phenomenological studies).
We will follow a different approach and estimate the fluctuations of the axial charge directly from the analytic expressions for the initial fields in Eq.~\nr{eq:fieldic}, including also the structure of fluctuations of the axial charge density in the transverse plane. Starting from the explicit form of the axial anomaly relation~\nr{eq:Anomaly} in Bjorken coordinates\footnote{Note that the transformation to co-moving coordinates can performed by expressing the left-hand side as $\nabla_{\mu} j^{\mu}=\frac{1}{\sqrt{-g}} \partial_{\mu} \Big( \sqrt{-g} j^{\mu}\Big)$, where $\nabla_{\mu}$ is the covariant (under coordinate transformations) derivative and the current $j^{\mu}$ transforms as contravariant vector. Similarly, the right hand side can be expressed as $\frac{1}{\sqrt{-g}} \epsilon^{\mu\nu\alpha\beta} F_{\mu\nu} F_{\alpha\beta}$, where the field strength $F_{\mu\nu}=\nabla_{\mu}A_{\nu}-\nabla_{\nu}A_{\mu}+ig[A_{\mu},A_{\nu}]$ and the Levi-Civita symbol $\frac{1}{\sqrt{-g}} \epsilon^{\mu\nu\alpha\beta}$ transform as covariant and contravariant tensors respectively.} and defining a shorter notation $\nudot(\xt) \equiv \tr E \cdot B$ for the divergence of the Chern-Simons current
\begin{eqnarray}
\left[\partial_{\tau}  + \frac{1}{\tau} \right] j^{\tau}_{(5)}(\xt)+\partial_{i}j^{i}_{(5)}(\xt)+\partial_{\eta} j^{\eta}_{(5)}(\xt)
=-\frac{g^2 N_f}{8\pi^2} \tr F_{\mu\nu}\tilde{F}^{\mu\nu}
= \frac{g^2 N_f}{2\pi^2}\nudot(\xt),
\end{eqnarray}
we first note that the term $\partial_{\eta} j^{\eta}_{(5)\xt}$ vanishes by virtue of the boost invariance assumption.  Next we note that -- at sufficiently early times -- we can neglect the effect of the axial currents $\partial_{i}j^{i}_{(5)}(\xt)$, such that the source term on the right-hand side
\begin{eqnarray}
\nudot(\xt)=\tr \Big[ E^{\eta}(\tau=0^{+},\xt) B^{\eta}(\tau=0^{+},\xt) \Big]+\mathcal{O}(\tau^2)
\end{eqnarray}
leads to local production of axial charge imbalance before axial charge starts to diffuse in the transverse plane. Based on this approximation one can then estimate the local density of axial charge per unit rapidity at each point in the transverse plane according to
\begin{eqnarray}
\label{eq:AxialChargeProduction}
\left.\frac{\ud N_5}{\ud^2\xt \ud\eta} \right|_{\tau\lesssim 1/\qs} \approx  \frac{\tau^2}{2} \frac{g^2 N_f}{2\pi^2}  \nudot(\xt,\tau=0^+)\;.
\end{eqnarray}
This allows us to compute axial charge production directly from the correlation functions of light-like Wilson lines. As we will discuss shortly the expectation value of the quantity $\nudot(\xt)$ is vanishes in accordance with the fact that there is no $CP$ violation in the process. Nevertheless, there can be sizeable fluctuations on an event-by-event basis, which are characterized by the correlation function $\langle\nudot(\xt)\nudot(\yt) \rangle$ at two different points $\xt,\yt$ in the transverse plane. Since $\nudot(\xt)$ is a dimension four operator, it is most naturally compared to the energy density $\varepsilon(\xt)$ of the system, and we will also compute the correlation functions of the energy-density $\langle\varepsilon(\xt)\varepsilon(\yt) \rangle$ for comparison.

\section{Energy density and Chern-Simons currents in the Glasma}
\label{sec:correlators}

Before we turn to the evaluation of correlation functions of the energy density and the divergence of the Chern-Simons current, we will briefly review the calculation of the corresponding one-point functions. Even though the results are well established in the literature~\cite{Lappi:2006hq,Chen:2015wia,Chen:2013ksa} this exercise is nevertheless useful to illustrate the procedure and fix our notations.

\subsection{Expectation values of one-point functions}
Based on the analytic expressions for the color-electric and color-magnetic fields at $\tau=0^{+}$ we can immediately compute the expectation value of the local energy density $\varepsilon(\xt)$ and the divergence of the Chern-Simons current $\nudot(\xt)$ as
\begin{eqnarray}
\langle \varepsilon(\xt) \rangle&=&\left\langle \tr\Big(E^{\eta}_{\xt} E^{\eta}_{\xt} + B^{\eta}_{\xt} B^{\eta}_{\xt}\Big) \right\rangle = (-ig)^2\Big(\delta^{ij}\delta^{kl} + \varepsilon^{ij}\varepsilon^{kl} \Big)\left\langle \tr\Big( [\alpha^{i}_{\xt},\beta^{j}_{\xt}],[\alpha^{k}_{\xt},\beta^{l}_{\xt}] \Big) \right\rangle\;, \\
\langle \nudot(\xt) \rangle&=& \qquad \left\langle \tr\Big(E^{\eta}_{\xt} B^{\eta}_{\xt}\Big)  \right \rangle\; \quad ~ = (-ig)^2 \qquad \delta^{ij}\varepsilon^{kl} \qquad ~~ \left\langle \tr\Big( [\alpha^{i}_{\xt},\beta^{j}_{\xt}],[\alpha^{k}_{\xt},\beta^{l}_{\xt}] \Big) \right\rangle.
\end{eqnarray}
Evaluating the color structures by decomposing $\alpha,\beta$ over the Lie Algebra, noting that  $\tr[t^{c} t^{c'}]=\frac{\delta^{cc'}}{2}$
and separating the averages over projectile and target fields, we obtain
\begin{eqnarray}
\langle \varepsilon(\xt) \rangle &=& (-ig)^2 \Big(\delta^{ij}\delta^{kl} + \varepsilon^{ij}\varepsilon^{kl} \Big) \frac{1}{2} if^{abc} if^{a'b'c}  \left\langle \alpha^{i,a}_{\xt} \alpha^{k,a'}_{\xt} \right\rangle  \left\langle \beta^{j,b}_{\xt} \beta^{l,b'}_{\xt} \right\rangle\;, \\
\langle \nudot(\xt) \rangle&=&(-ig)^2 \qquad \delta^{ij}\varepsilon^{kl} \qquad ~~ \frac{1}{2} if^{abc} if^{a'b'c}  \left\langle \alpha^{i,a}_{\xt} \alpha^{k,a'}_{\xt} \right\rangle  \left\langle \beta^{j,b}_{\xt} \beta^{l,b'}_{\xt} \right\rangle\;.
\end{eqnarray}
Since only color-singlet expectation values are non-vanishing, such that 
\begin{eqnarray}
\langle \alpha^{i,a}_{\xt} \alpha^{k,b}_{\yt} \rangle = W^{ik}_{(U)}(\xt,\yt)~\delta^{ab} \; ,
\end{eqnarray}
we can evaluate the color structure as $f^{abc}f^{abc}=\nc (\nc^2-1)$. Upon factorization of the averages of the projectile and target fields we obtain
\begin{eqnarray}
(-ig)^2 \left\langle \tr\Big( [\alpha^{i}_{\xt},\beta^{j}_{\xt}][\alpha^{k}_{\xt},\beta^{l}_{\xt}] \Big) \right\rangle =  g^2\frac{\nc (\nc^2-1)}{2}~W^{ik}_{(U)}(\xt,\xt)~W^{jl}_{(V)}(\xt,\xt)
\end{eqnarray}
where $W^{ik}_{(U/V)}(\xt\yt)$ are the Weizs\"acker-Williams gluon distributions of the two nuclei
\begin{eqnarray}
W^{ik}_{(U)}(\xt,\yt)=\frac{1}{\nc^2-1} \langle \alpha^{i,a}_{\xt} \alpha^{k,a}_{\yt}  \rangle \; .
\end{eqnarray}
Generally speaking, the Weizs\"acker-Williams distribution can be further decomposed into various different tensor structures. We start from  the usual momentum space decomposition into unpolarized $G^{(1)}$ and linearly polarized $h^{(1)}_{\bot}$ gluon distributions in an unpolarized hadron\footnote{Note that taking into account impact parameter $\bt$ dependence the most general decomposition requires additional tensor  structures involving $\bt$ as well as combinations of $\bt$ and $\kt$. However, since we are only interested in the application to the collisions of large nuclei, we will ignore these subtleties and proceed as usual. We refer the interested reader to Ref.~\cite{Chen:2013ksa,Chen:2015wia} for a detailed discussion on the implications for energy density correlators in the glasma. 
}
\footnote{Normalization conventions for the Weizs\"acker-Williams distributions vary, for example the one introduced in~\cite{Marquet:2016cgx} is related to the one here by
$$
\mathcal{F}^{(3)}_{gg}(\kt) = \frac{\nc^2-1}{4\pi^3}\int \ud^2\bt \; G^{(1)}(\bt,\ktt).
$$
}
\begin{eqnarray}
\widetilde{W}^{ij}(\bt,\kt)=\frac{1}{2}\delta^{ij} G^{(1)}(\bt,\ktt) - \frac{1}{2} \left( \delta^{ij}-2 \frac{\kt^{i}\kt^{j}}{\kt^2}\right) h^{(1)}_{\bot}(\bt,\ktt) 
\end{eqnarray}
where $\widetilde{W}^{ij}(\bt,\kt)= \int \ud^2 \rt ~W^{ij}_{(U)}(\bt+\rt/2,\bt-\rt/2)~e^{-i\kt\rt}$. The corresponding tensor decomposition in coordinate space takes the form 
\begin{eqnarray}
W^{ij}_{(U)}(\xt,\yt)=\frac{1}{2}\delta^{ij} G^{(1)}(\xt,\yt) + \frac{1}{2} \left( \delta^{ij}-2 \frac{(\xt-\yt)^{i}(\xt-\yt)^{j}}{|\xt-\yt|^2}\right) h^{(1)}_{\bot}(\xt,\yt)
\end{eqnarray}
where the coordinate space functions $G^{(1)}(\xt,\yt)$ and $h^{(1)}_{\bot}(\xt,\yt)$ are given by
\begin{eqnarray}
G^{(1)}(\xt,\yt) &=& \frac{1}{2\pi} \int \ud \ktt \ktt~J_{0}(\kt |\xt-\yt|)~G^{(1)}\Big(\frac{\xt+\yt}{2},\ktt\Big)\;, \\
h^{(1)}_{\bot}(\xt,\yt)&=& \frac{1}{2\pi} \int \ud\ktt \ktt~J_{2}(\kt |\xt-\yt|)~h^{(1)}_{\bot}\Big(\frac{\xt+\yt}{2},\ktt\Big)\;.
\end{eqnarray}
Note that due to the angular structure of the integration, $h^{(1)}_{\bot}(\xt,\yt)$ is {\it not} simply the Fourier transform of the linearly polarized gluon distribution $h^{(1)}_{\bot}\left(\bt,\ktt\right),$ but involves a Bessel function of order two.

The Weizs\"acker-Williams distribution $W^{ij}$ is a $2\times2$ matrix with eigenvalues $G^{(1)} \pm h^{(1)}_{\bot}$. As expectation values of positive definite operators, both $W^{ij}_{(U)}(\xt,\xt)$ (at the same coordinate $\xt=\yt$) and the impact parameter averaged $\widetilde{W}^{ij}(\kt)$ (for general $\kt$) should be positive definite.  This leads to positivity constraints~\cite{Mulders:2000sh} in both coordinate and momentum space, which in our notation read $G^{(1)}(\xt,\xt) \geq |h^{(1)}_{\bot}(\xt,\xt)|$ and $G^{(1)}(\kt) \geq |h^{(1)}_{\bot}(\kt)|$. Note that even if the momentum space distributions saturate the positivity bound ($G^{(1)}(\ktt) = h^{(1)}_{\bot}(\ktt)$ in our normalization) as is expected at high transverse momentum, this is not true for the  coordinate space functions due to the behavior of the Bessel functions near the origin, which will be important for our discussion in the following. 

Collecting everything and expressing the result in terms of the $G^{(1)}$ and $h^{(1)}$ we obtain the following expression for the local operator expectation values
\begin{eqnarray}
\langle \varepsilon(\xt) \rangle=\frac{g^2~\nc (\nc^2-1)}{2}~G^{(1)}_{(U)}(\xt,\xt)~G^{(1)}_{(V)}(\xt,\xt) \;, \qquad 
\langle \nudot(\xt) \rangle=0\;, 
\end{eqnarray}
where the index contractions lead to a vanishing result for the expectation value of the $CP$ odd operator $\nudot$. We see that the linearly polarized distribution does not contribute to the expectation value\footnote{Note that the Ref.~\cite{Lappi:2006hq} assumes the case of a full linear polarization $G^{(1)}(\kt) = h^{(1)}(\kt)$, which is true in the MV model at large transverse momentum (see the discussion in Sec.~\ref{sec:gg}). Our present result shows that this simplifying assumption did not affect the final result.}.

\subsection{Saturation models for Weiszs\"{a}cker-Williams distribution}
\label{sec:dipmodels}
In order to provide explicit results for the one- and two-point correlation functions, we need to specify a model for the Weiszs\"{a}cker-Williams gluon distribution. We follow previous works and exploit the fact~\cite{Kharzeev:2003wz,Jalilian-Marian:1997xn} that in Gaussian models the Weiszs\"{a}cker-Williams gluon distribution can be related to the Dipole gluon distribution for which a number of phenomenologically useful parametrizations exist. Based on this standard procedure, described for completeness in Appendix \ref{app:EvaluationGaussianModel}, we obtain
\begin{eqnarray}
W^{ik}_{(U)}(\xt,\yt)=\frac{1}{g^2\nc} \left(\frac{\partial^{i}_{\xt} \partial^{k}_{\yt} \ln(D^{(U)}_{\xt\yt}) }{\ln(D^{(U)}_{\xt\yt})} \right) \left( \Big(D^{(U)}_{\xt\yt}\Big)^\frac{2 \nc^2}{\nc^2-1} -1 \right)\;.
\end{eqnarray}
where $D^{(U)}_{\xt\yt}=\frac{1}{\nc} \left\langle \tr \Big(U_{\xt} U^{\dagger}_{\yt}\Big) \right\rangle$ is the expectation value of the dipole operator. 
From this it is relatively easy, assuming that the dipole distribution only depends on the distance $r\equiv |\xt-\yt|$, to extract the individual distributions as
\begin{eqnarray}
 G^{(1)}(r) &=& 
\frac{1}{g^2\nc}
 \frac{1-  \Big(D^{(U)}(r)\Big)^{\frac{2 \nc^2}{\nc^2-1} } }{\ln \left(D^{(U)}(r)\right)}
\left(\partial_r^2+ \frac{1}{r}\partial_r\right)
\ln\left( D^{(U)}(r) \right)
 \\
h^{(1)}_{\bot} (r) &=& 
\frac{1}{g^2\nc}
 \frac{1- \Big(D^{(U)}(r)\Big)^{\frac{2 \nc^2}{\nc^2-1} } }{\ln \left(D^{(U)}(r)\right)}
\left(\partial_r^2- \frac{1}{r}\partial_r\right)
\ln\left( D^{(U)}(r) \right).
 \end{eqnarray}

We can explicitly evaluate the correlation function in a number of simple models. One finds for instance that in the Golec-Biernat Wusthoff (GBW) model~\cite{Golec-Biernat:1998js} for the dipole amplitude
\begin{eqnarray} 
D^{\rm GBW}_{\xt\yt}=\exp\left(-\frac{\qs^2}{4}  (\xt-\yt)^2\right)
\end{eqnarray}
the linearly polarized gluon distribution vanishes identically $h^{(1)}_{\bot,{\rm GBW}}(\xt,\yt)=0$ and the un-polarized gluon distribution is simply given by
\begin{eqnarray}
G^{(1)}_{\rm GBW}(\xt,\yt)=\frac{Q_{s}^{2}}{g^2 \nc}~\frac{1-e^{-\frac{2\nc^2}{\nc^2-1}\frac{\qs^2}{4} (\xt-\yt)^2}}{\frac{\qs^2}{4}(\xt-\yt)^2}\;.
\end{eqnarray}
such that in the limit $\yt \to \xt$, relevant for the expectation value of the local energy density
\begin{eqnarray}\label{eq:gbwdistzero}
G^{(1)}_{\rm GBW}(\xt,\xt)= \frac{\qs^2}{g^2 \cf}\;.
\end{eqnarray}
Conversely, in the (screened) McLerran-Venugopalan (MV) model the dipole amplitude is given by
\begin{eqnarray}
\label{eq:DipMV}
D^{\rm MV}_{\xt\yt}=\exp\left(\frac{g^4\mu^2_{0}}{4\pi m^2} \Big(m|\xt-\yt| K_{1}(m|\xt-\yt|)-1\Big) \right),
\end{eqnarray} 
such that the unpolarized and linearly polarized distributions become
\begin{eqnarray}
\label{eq:G1MV}
&&G^{(1)}_{\rm MV}(\xt,\yt)=\\
&&+\frac{g^4\mu^2_{0}}{4\pi g^2\nc} \Big(m|\xt-\yt| K_{1}(m|\xt-\yt|)-2K_{0}(m|\xt-\yt|)\Big) \frac{1-e^{\frac{2\nc^2}{\nc^2-1}\frac{g^4\mu^2_{0}}{4\pi m^2} \Big(m|\xt-\yt| K_{1}(m|\xt-\yt|)-1\Big)}}{\frac{g^4\mu^2_{0}}{4\pi m^2} \Big(m|\xt-\yt| K_{1}(m|\xt-\yt|)-1\Big) }\;. \nonumber \\
\label{eq:h1MV}
&&h^{(1)}_{\bot,\rm MV}(\xt,\yt)=\\
&&-\frac{g^4\mu^2_{0}}{4\pi g^2\nc}~m|\xt-\yt| K_{1}(m|\xt-\yt|)~\frac{1-e^{\frac{2\nc^2}{\nc^2-1}\frac{g^4\mu^2_{0}}{4\pi m^2} \Big(m|\xt-\yt| K_{1}(m|\xt-\yt|)-1\Big)}}{\frac{g^4\mu^2_{0}}{4\pi m^2} \Big(m|\xt-\yt| K_{1}(m|\xt-\yt|)-1\Big) }\;. \nonumber 
\end{eqnarray}
At fixed coupling $g$ the function $G^{(1)}_{\rm MV}(\xt,\yt)$ is logarithmically divergent in the limit $\yt \to \xt$ 
\begin{equation}
\lim_{\yt \to \xt}G^{(1)}_{\rm MV}(\xt,\yt)=-\frac{g^4\mu^2_{0} \nc}{2\pi g^2(\nc^2-1)} \left(1+2\gamma_{E} +\ln\Big(\frac{m^2|\xt-\yt|^2}{4}\Big)\right)\;, 
\end{equation}
whereas $h^{(1)}_{\bot,\rm MV}(\xt,\yt)$ remains finite
\begin{eqnarray}
\lim_{\yt \to \xt}h^{(1)}_{\bot,\rm MV}(\xt,\yt)&=&+\frac{g^4\mu^2_{0} \nc}{2\pi g^2(\nc^2-1)}\;.
\end{eqnarray}
We will take the suggestion of some previous works (e.g.~\cite{Iancu:2015joa}) and regulate the logarithmic divergence by introducing a running of the coupling via the following replacement in Eqs.~\nr{eq:G1MV},~\nr{eq:h1MV}:
\begin{eqnarray}
g^{4}\mu^2_{0} \to  g^2\Big(\mu^2\Big)~g^2\Big(\frac{1}{|\xt-\yt|^2}\Big)~\mu^2\;, \qquad 
\end{eqnarray}
with the coordinate space running coupling
\begin{eqnarray}
g^2\Big(\frac{1}{|\xt-\yt|^2}\Big)=g^2(\mu^2)~\frac{\ln\Big(\frac{\mu^2}{\Lambda^2}\Big)}{\ln\Big(\frac{4 e^{-2\gamma_e}}{\Lambda^2 |\xt-\yt|^2}+e \Big)}\;.
\end{eqnarray}
We then absorb the superfluous parameters into physical ones by expressing the correlators in terms of the physical momentum scale of the problem, the saturation scale $\qs$. This can be done by taking the limit 
$\Lambda \sim m \ll \mu \sim \qs$, in which the unpolarized distribution becomes 
\begin{equation}
\lim_{\yt \to \xt}G^{(1)}_{\rm MV}(\xt,\yt)
\approx  \frac{g^2(\mu^2) \mu^2  }{4\pi \cf} 
\ln\Big(\frac{\mu^2}{m^2}\Big).
\end{equation}
We then require that this has the same normalization in term of the saturation scale $\qs$ as in the GBW parametrization, see Eq.\nr{eq:gbwdistzero}. This can be achieved by setting
$$
\qs^2=  \frac{g^4(\mu^2)~\mu^2}{4\pi} 
\ln\Big(\frac{\mu^2}{m^2}\Big).
$$
Expressed in terms of $\qs$ we now have the same short distance behavior as in the GBW model
\begin{eqnarray}
\lim_{\yt \to \xt}G^{(1)}_{\rm MV}(\xt,\yt)=\frac{\qs^2}{g^2 \cf}\;,  \qquad \lim_{\yt \to \xt}h^{(1)}_{\bot,\rm MV}(\xt,\yt)=0\;, 
\end{eqnarray}
and we will employ this prescription  in the following when presenting numerical results.

It is interesting to note that a for a dipole parametrization that has an UV anomalous dimension, i.e. $\ln D_{\xt\yt} \sim -(\xt-\yt)^{2\gamma}$, corresponding to $h^{(1)}_{\bot}(\ktt) \sim G^{(1)}(\ktt)\sim \ktt^{-2\gamma}$, the limiting behavior at small distance is given by $\lim_{\yt\to\xt}\frac{h^{(1)}_{\bot}(\xt,\yt)}{G^{(1)}(\xt,\yt)} = \frac{1-\gamma}{\gamma}$. 
Fits to HERA data using the BK equation favor values  $\gamma \gtrsim 1$ for the initial condition, which evolves to $\gamma \lesssim 1$ during the evolution. All of these are in the region $\gamma \geq 1/2$ required by the positivity bound $G^{(1)}(\xt,\xt)\geq h^{(1)}_{\bot}(\xt,\xt)$. At the limiting value $\gamma=1$ of the MV model the analytical structure changes: $h^{(1)}_{\bot}$ changes sign and for $\gamma\geq 1$ the convergence of the Fourier-integral for the coordinate space  distribution $G^{(1)}$ in terms of the momentum space one starts to require regularization; see the related discussion in \cite{Giraud:2016lgg}.

\subsection{Expectation values of two-point correlation functions}
\label{sec:correlatorcalculation}

We now turn to the evaluation of correlation functions of $\varepsilon(\xt)\varepsilon(\yt)$ and $\nudot(\xt)\nudot(\yt)$ characterizing local fluctuations of the energy density and divergence of the Chern-Simons current in the transverse plane. By performing the same steps as outlined above, we obtain for the correlation functions
\begin{eqnarray}
\varepsilon(\xt)\varepsilon(\yt)&=& \left\langle \tr\Big(E^{\eta}_{\xt} E^{\eta}_{\xt} + B^{\eta}_{\xt} B^{\eta}_{\xt}\Big)~\tr\Big(E^{\eta}_{\yt} E^{\eta}_{\yt} + B^{\eta}_{\yt} B^{\eta}_{\yt}\Big) \right\rangle,
\\ 
\nudot(\xt)\nudot(\yt)&=& \left\langle \tr\Big(E^{\eta}_{\xt} B^{\eta}_{\xt}\Big)~\tr\Big(E^{\eta}_{\yt}B^{\eta}_{\yt}\Big) \right\rangle\;, \nonumber \\
\end{eqnarray}
the following expressions
\begin{equation} \label{eq:corr}
\begin{split}
\varepsilon(\xt)\varepsilon(\yt)= (-ig)^4~\Big(\delta^{ij}\delta^{kl} + \varepsilon^{ij}\varepsilon^{kl} \Big)
&
\Big(\delta^{i'j'}\delta^{k'l'} + \varepsilon^{i'j'}\varepsilon^{k'l'} \Big)  
\\ &   \times
\left\langle \tr\Big( [\alpha^{i}_{\xt},\beta^{j}_{\xt}] \ [\alpha^{k}_{\xt},\beta^{l}_{\xt}] \Big)  \tr\Big( [\alpha^{i'}_{\yt},\beta^{j'}_{\yt}] \ [\alpha^{k'}_{\yt},\beta^{l'}_{\yt}] \Big) \right\rangle  
\\
\nudot(\xt)\nudot(\yt)= (-ig)^4~~\qquad \delta^{ij}\varepsilon^{kl} \qquad & ~\quad 
 \qquad \delta^{i'j'}\varepsilon^{k'l'}  
\\  &    \times
\left\langle \tr\Big( [\alpha^{i}_{\xt},\beta^{j}_{\xt}] \ [\alpha^{k}_{\xt},\beta^{l}_{\xt}] \Big)  \tr\Big( [\alpha^{i'}_{\yt},\beta^{j'}_{\yt}] \ [\alpha^{k'}_{\yt},\beta^{l'}_{\yt}] \Big) \right\rangle .
\end{split}
\end{equation}
We now have to evaluate correlation functions of the gluon field
\begin{eqnarray}
&&\left\langle \tr\Big( [\alpha^{i}_{\xt},\beta^{j}_{\xt}] \ [\alpha^{k}_{\xt},\beta^{l}_{\xt}] \Big)  \tr\Big( [\alpha^{i'}_{\yt},\beta^{j'}_{\yt}] \ [\alpha^{k'}_{\yt},\beta^{l'}_{\yt}] \Big) \right\rangle = \\
&& \qquad \qquad \qquad \frac{1}{4}~if^{abe}~if^{cde}~if^{a'b'e'}~if^{c'd'e'}~\langle \alpha^{i,a}_{\xt} \alpha^{k,c}_{\xt} \alpha^{i',a'}_{\yt} \alpha^{k',c'}_{\yt} \rangle \langle \beta^{j,b}_{\xt} \beta^{l,d}_{\xt} \beta^{j',b'}_{\yt} \beta^{l',d'}_{\yt}  \rangle\;. \nonumber
\end{eqnarray}
Even though it is in principle possible to evaluate such objects numerically in Gaussian models as discussed e.g. in~\cite{Dusling:2017dqg}, we will follow a different approach in order to obtain some analytic insight. Namely, we will assume that the four-point correlation functions of the gluon fields can be factorized into products of two-point correlation functions such that
\begin{eqnarray}
\label{eq:DoublePartonFactorization}
\langle \alpha^{i,a}_{\xt} \alpha^{k,c}_{\xt} \alpha^{i',a'}_{\yt} \alpha^{k',c'}_{\yt} \rangle &=& 
 \overbrace{\delta^{ac} \delta^{a'c'} W_{(U)}^{ik}(\xt,\xt)W^{i'k'}(\yt,\yt)}^{\text{disconnected}}   
 \\ \nonumber && \qquad
 + 
\underbrace{\delta^{aa'} \delta^{cc'} W_{(U)}^{ii'}(\xt,\yt)  W_{(U)}^{kk'}(\xt,\yt) + \delta^{ac'} \delta^{ca'} W_{(U)}^{ik'}(\xt,\yt) W_{(U)}^{ki'}(\xt,\yt)}_{\text{connected}} \;. \nonumber 
\end{eqnarray}
and similarly for the second nucleus
\begin{eqnarray}
\label{eq:DoublePartonFactorization2}
\langle \beta^{j,b}_{\xt} \beta^{l,d}_{\xt} \beta^{j',b'}_{\yt} \beta^{l',d'}_{\yt}  \rangle &=& 
  \overbrace{\delta^{bd} \delta^{b'd'} W^{jl}_{(V)}(\xt,\xt) W^{j'l'}_{(V)}(\yt,\yt)}^{\text{disconnected}} 
 \\ \nonumber && \qquad  
  + 
\underbrace{\delta^{bb'} \delta^{dd'} W^{jj'}_{(V)}(\xt,\yt) W^{ll'}_{(V)}(\xt,\yt)  + \delta^{bd'} \delta^{db'} W^{jl'}_{(V)}(\xt,\yt)  W^{lj'}_{(V)}(\xt,\yt)}_{\text{connected}}\;. \nonumber 
\end{eqnarray}
Stated differently, this procedure corresponds to a factorization of the relevant double parton distribution into all possible products of single parton distributions. We note that this approximation scheme has been frequently employed in the literature e.g. in the context of di-hadron correlations (Glasma graphs)~\cite{Dumitru:2008wn,Gelis:2009wh,Dusling:2009ni,Dumitru:2010iy,Dusling:2012iga,Dusling:2012wy,Dusling:2013qoz} and the quality of approximation has been investigated e.g. in~\cite{Lappi:2015vta}.

Based on the above expression for the four-point correlation functions of the gluon fields, we can then proceed to evaluate the color structures in the expressions. Distinguishing the terms by connected and disconnected contractions as indicated in Eqs.~(\ref{eq:DoublePartonFactorization}), (\ref{eq:DoublePartonFactorization2}), the relevant color factors are given by
\begin{align}
&\text{disconnected-disconnected: } \nonumber \\
&\quad if^{abe}~if^{cde}~if^{a'b'e'}~if^{c'd'e'} \delta^{ac} \delta^{a'c'} \delta^{bd} \delta^{b'd'}=f^{abe}f^{abe}f^{a'b'e'}f^{a'b'e'}=\nc^2 (\nc^2-1)^2\;, \\
&\text{disconnected-connected: } \nonumber \\
&\quad if^{abe}~if^{cde}~if^{a'b'e'}~if^{c'd'e'} \delta^{ac} \delta^{a'c'} \delta^{bb'} \delta^{dd'}=f^{abe}f^{ade}f^{a'be'}f^{a'de'}=\nc^2 (\nc^2-1)\;, \\
& \quad if^{abe}~if^{cde}~if^{a'b'e'}~if^{c'd'e'} \delta^{ac} \delta^{a'c'} \delta^{bd'} \delta^{db'}=f^{abe}f^{ade}f^{a'de'}f^{a'be'}=\nc^2 (\nc^2-1)\;, \\
& \text{connected-disconnected: }\nonumber \\
& \quad  if^{abe}~if^{cde}~if^{a'b'e'}~if^{c'd'e'}  \delta^{aa'} \delta^{cc'} \delta^{bd} \delta^{b'd'}=f^{abe}f^{cbe}f^{ab'e'}f^{cb'e'}=\nc^2(\nc^2-1)\;, \\
& \quad if^{abe}~if^{cde}~if^{a'b'e'}~if^{c'd'e'}  \delta^{ac'} \delta^{ca'} \delta^{bd} \delta^{b'd'}=f^{abe}f^{cbe}f^{cb'e'}f^{ab'e'}=\nc^2(\nc^2-1)\;, \\
& \text{connected-connected: }\nonumber \\
& \quad if^{abe}~if^{cde}~if^{a'b'e'}~if^{c'd'e'}  \delta^{aa'} \delta^{cc'} \delta^{bb'} \delta^{dd'}=f^{abe}f^{cde} f^{abe'}f^{cde'}=\nc^2(\nc^2-1)\;, \\
& \quad if^{abe}~if^{cde}~if^{a'b'e'}~if^{c'd'e'}  \delta^{aa'} \delta^{cc'} \delta^{bd'} \delta^{db'}=f^{abe}f^{cde} f^{ade'}f^{cbe'}=\frac{1}{2}\nc^2(\nc^2-1)\;, \\
& \quad if^{abe}~if^{cde}~if^{a'b'e'}~if^{c'd'e'}  \delta^{ac'} \delta^{ca'} \delta^{bb'} \delta^{dd'}=f^{abe}f^{cde} f^{cbe'}f^{ade'}=\frac{1}{2}\nc^2(\nc^2-1)\;, \\
& \quad if^{abe}~if^{cde}~if^{a'b'e'}~if^{c'd'e'}  \delta^{ac'} \delta^{ca'} \delta^{bd'} \delta^{db'}=f^{abe}f^{cde} f^{cde'}f^{abe'}=\nc^2(\nc^2-1)\;,
\end{align}
where we used the identities $\tr\Big[ T^{a}_{adj} T^{b}_{adj}\Big]=\nc~\delta^{ab}$ and $\tr\Big[ T^{a}_{adj} T^{b}_{adj} T^{a}_{adj} T^{c}_{adj}\Big]=\frac{1}{2} \nc^2 \delta^{bc} $ to evaluate the final expressions. Collecting all the different terms we then obtain for the correlation function
\begin{align}
\label{eq:ww4pt}
&\left\langle \tr\Big( [\alpha^{i}_{\xt},\beta^{j}_{\xt}],[\alpha^{k}_{\xt},\beta^{l}_{\xt}] \Big)  \tr\Big( [\alpha^{i'}_{\yt},\beta^{j'}_{\yt}],[\alpha^{k'}_{\yt},\beta^{l'}_{\yt}] \Big) \right\rangle = \\
& \qquad +\frac{\nc^2 (\nc^2-1)^2}{4} \left[ W_{(U)}^{ik}(\xt,\xt)W^{i'k'}(\yt,\yt) W^{jl}_{(V)}(\xt,\xt) W^{j'l'}_{(V)}(\yt,\yt)  \right]  \nonumber \\
& \qquad  +\frac{\nc^2 (\nc^2-1)}{4} ~\left[ W_{(U)}^{ik}(\xt,\xt)W^{i'k'}(\yt,\yt) \Big(W^{jj'}_{(V)}(\xt,\yt) W^{ll'}_{(V)}(\xt,\yt)  + W^{jl'}_{(V)}(\xt,\yt)  W^{lj'}_{(V)}(\xt,\yt) \Big)\right]  \nonumber \\
& \qquad +\frac{\nc^2 (\nc^2-1)}{4}~\left[  \Big(W_{(U)}^{ii'}(\xt,\yt)  W_{(U)}^{kk'}(\xt,\yt) + W_{(U)}^{ik'}(\xt,\yt) W_{(U)}^{ki'}(\xt,\yt) \Big) W^{jl}_{(V)}(\xt,\xt) W^{j'l'}_{(V)}(\yt,\yt)\right] \nonumber \\
& \qquad +\frac{\nc^2 (\nc^2-1)}{4}~\left[ W_{(U)}^{ii'}(\xt,\yt) W_{(U)}^{kk'}(\xt,\yt) \Big( W^{jj'}_{(V)}(\xt,\yt) W^{ll'}_{(V)}(\xt,\yt)+ \frac{1}{2} W^{jl'}_{(V)}(\xt,\yt)  W^{lj'}_{(V)}(\xt,\yt)\Big)\right. \nonumber \\
&\qquad \qquad \qquad \qquad \left.+W_{(U)}^{ik'}(\xt,\yt) W_{(U)}^{ki'}(\xt,\yt)\Big(\frac{1}{2}W^{jj'}_{(V)}(\xt,\yt) W^{ll'}_{(V)}(\xt,\yt)+W^{jl'}_{(V)}(\xt,\yt)  W^{lj'}_{(V)}(\xt,\yt)\Big)\right] \nonumber
\end{align}
where we note that  -- as usual -- all terms involving a connected contraction are suppressed by a factor $1/(\nc^2-1)$ relative to the fully disconnected contribution.

By performing also all of the contractions of the transverse tensors, we obtain after some algebra our result for the two point correlation function of the energy density
\begin{align}
\label{eq:epsepscorr}
&\langle\varepsilon(\xt)\varepsilon(\yt)\rangle - \langle\varepsilon(\xt)\rangle   \langle\varepsilon(\yt)\rangle=  \\
&\qquad \frac{g^4\nc^2 (\nc^2-1)}{4} G^{(1)}_{(U)}(\xt,\xt) G^{(1)}_{(U)}(\yt,\yt) \left[ \Big(G^{(1)}_{(V)}(\xt,\yt)\Big)^2 + \Big(h^{(1)}_{\bot (V)}(\xt,\yt)\Big)^2 \right] \nonumber \\
&\qquad + \frac{g^4\nc^2 (\nc^2-1)}{4} G^{(1)}_{(V)}(\xt,\xt) G^{(1)}_{(V)}(\yt,\yt) \left[\Big(G^{(1)}_{(U)}(\xt,\yt)\Big)^2 + \Big(h^{(1)}_{\bot (U)}(\xt,\yt)\Big)^2 \right]   \nonumber  \\
&\qquad + \frac{g^4\nc^2 (\nc^2-1)}{4} \left[\Big(G^{(1)}_{(U)}(\xt,\yt)\Big)^2 \Big(G^{(1)}_{(V)}(\xt,\yt)\Big)^2 + \Big(h^{(1)}_{\bot(U)}(\xt,\yt)\Big)^2 \Big(h^{(1)}_{\bot(V)}(\xt,\yt)\Big)^2 \right] \nonumber  \\
&\qquad + \frac{g^4\nc^2 (\nc^2-1)}{8} \left[\Big(G^{(1)}_{(U)}(\xt,\yt)\Big)^2 \Big(h^{(1)}_{\bot(V)}(\xt,\yt)\Big)^2   +  \Big(h^{(1)}_{\bot(U)}(\xt,\yt)\Big)^2 \Big(G^{(1)}_{(V)}(\xt,\yt)\Big)^2 \right]\;. \nonumber
\end{align}
which receives three distinct contributions, related to the disconnected-connected, connected-disconnected and connected-connected contributions\footnote{We note that the above expression corrects the earlier result of \cite{Muller:2011bb}, where the connected-connected term in the last two lines was given incorrectly. We have checked the calculation of Ref.~\cite{Muller:2011bb} step-by-step. In the notation of the reference, we find that the pre-factor of the fully connected contribution to $M_{1}$ should be $3/16$ instead of $3/8$ and that  $M_{5}+M_{6}+M_{8}+M_{9}=\frac{g^4}{16}\nc^2 (\nc^2-1)[G^2(\xt-\yt)-E^2(\xt-\yt)-F^2(\xt-\yt)]$ featuring a relative minus sign between the unpolarized and linearly polarized contributions.}. In contrast for the Chern-Simons correlator, all disconnected contractions vanish identically and only the connected-connected contractions give rise to a non-vanishing contribution. Our final result and the central result of this paper reads
\begin{eqnarray}
\label{eq:nunucorr}
&&\langle \nudot(\xt) \nudot(\yt) \rangle= \\
&&\qquad \qquad \frac{3g^4\nc^2 (\nc^2-1)}{32}  \left[\Big(G^{(1)}_{(U)}(\xt,\yt)\Big)^2 \Big(G^{(1)}_{(V)}(\xt,\yt)\Big)^2 - \Big(h^{(1)}_{\bot(U)}(\xt,\yt)\Big)^2 \Big(h^{(1)}_{\bot(V)}(\xt,\yt)\Big)^2 \right]\;. \nonumber 
\end{eqnarray}
We note that the unpolarized and linearly polarized distributions contribute with different relative signs. However, as discussed in Sec.~\ref{sec:dipmodels}, at sufficiently small distances $|\xt-\yt|$ the unpolarized contribution dominates and the correlation function is manifestly positive. Of course, the calculation outlined above can also be performed more or less entirely using modern computer algebra tools such as \textsc{FeynCalc}~\cite{Mertig:1990an,Shtabovenko:2016sxi} or \textsc{Form}~\cite{Vermaseren:2000nd} and we have cross-checked our results in this way.

\section{Diagrammatic analysis in momentum space}
\label{sec:momspace}
\subsection{Correlators in momentum space}

Even though we have performed the entire calculation above in coordinate space, our calculation is in fact closely related to the Glasma graph analysis of double inclusive particle production. In order to illustrate the similarities and differences it is useful to generalize our previous expressions to finite time by approximating the dynamics in the forward light-cone $(\tau>0)$ in terms of the free field evolution. Based on the linearized evolution equations for abelian gauge fields
\begin{equation}\label{eq:lineom}
\partial_{\tau}\frac{1}{\tau} \partial_{\tau} \tau^2A^{\eta}= \tau \nabt^2 A^{\eta}\;, \qquad \qquad
\partial_{\tau} \tau \partial_{\tau} A_{i} = \tau \Big( \delta^{ij} \nabt^2   - \partial_{i} \partial_{j} \Big) A_{j}
\end{equation} 
the dynamics of the two independent polarizations corresponding to non-zero $E^{\eta}$ and respectively $B^{\eta}$ at $\tau=0^{+}$ decouples from each other. By matching the general solution of Eq.~(\ref{eq:lineom}) in Fourier space $A_\mu(\kt) =  \int \ud^2 \xt \; e^{-i\kt\cdot \xt} A_\mu(\xt) $ to the relevant initial conditions in Eqs.~(\ref{eq:potic}), (\ref{eq:fieldic}), we can immediately obtain a solution of the form (c.f. \cite{McLerran:2016snu}) \footnote{By construction this solution satisfies the Coloumb type gauge condition $\kt \cdot \bf{A}(\tau,\kt)=0$. It is thus a gauge transformation of the usual initial gauge potentials~\nr{eq:potic}.
}
\begin{eqnarray}                         
A_{\eta}(\tau,\kt) &=& -\tau^2 A^\eta  = \frac{\tau}{\ktt} E^{\eta}(\tau=0^{+},\kt) J_{1}(\ktt \tau)\;, \\
A_{i}(\tau,\kt) &=& A_{i}(0,\kt) J_{0}(\ktt \tau) = -i \varepsilon^{ij} \frac{\kt^j}{\ktt^2}~B^{\eta}(\tau=0^{+},\kt)~J_{0}(\ktt \tau)\;.
\end{eqnarray}
Staying consistently at lowest order in the abelian approximation to the dynamics in the forward light-cone, the non-abelian field strength can be determined as
\begin{align}
E^{\eta}(\tau,\kt)&=~~~\frac{1}{\tau}\partial_{\tau} A_{\eta}(\tau,\kt)~~~~~ = ~~~~~~~~~~~~E^{\eta}(\tau=0^{+},\kt)~J_{0}(\ktt \tau)\;, \\
E^{i}(\tau,\kt)&= ~~~ \partial_{\tau} A_{i}(\tau,\kt) ~~~~~~~~= -i \varepsilon^{ij} \frac{\kt^j}{\ktt}~B^{\eta}(\tau=0^{+},\kt)~J_{1}(\ktt \tau)\;, \\
B^{\eta}(\tau,\kt)&=~ -i\varepsilon^{ij}~\kt^{i}~A_{j}(\tau,\kt) = ~~~~~~~~~~~~B^{\eta}(\tau=0^{+},\kt)~J_{0}(\ktt \tau)\;, \\
B^{i}(\tau,\kt) &=  -i\varepsilon^{ij}~\kt^{j} A_{\eta}(\tau,\kt)/\tau = -i \varepsilon^{ij}~\frac{\kt^{j}}{\ktt}~E^{\eta}(\tau=0^{+},\kt) J_{1}(\ktt \tau)\;.
\end{align}

One subtle issue is that the quality of the abelian approximation for the dynamics in the forward light-cone depends on the gauge choice. Even though the above expressions show that dynamics in the abelian approximation can be entirely formulated in terms of correlation functions of chromo-electric and chromo-magnetic fields, objects such as $E^{a}(\tau=0^{+},\xt)E^{b}(\tau=0^{+},\yt)$ are in fact not gauge invariant. One natural gauge choice is the transverse Coulomb gauge $\partial_{i}A^{i}(\tau=0^{+},\xt)=0$ which minimizes the transverse gauge field amplitudes, and it has been established from numerical simulations in~\cite{Blaizot:2010kh} that the effects of final state interactions at $\tau>0$ become small in this gauge.

It is not generally known how to find the gauge transformation to Coulomb gauge analytically. However, the problem becomes considerably simpler in the case where either the projectile or target can be considered as dilute~\cite{Dumitru:2001ux,Blaizot:2008yb}. Specifically, if this is the case for the second nucleus ($V_{\xt}=1+ig\mathcal{A}_{(V)}(\xt)$), a gauge transformation with $V^{\dagger} U^{\dagger}$ yields the desired result to leading order in the dilute expansion. One finds that in this case, the non-vanishing components of the field strength tensor are given by 
\begin{eqnarray}
\left. E^{\eta}(\tau=0^{+},\xt) \right|_{Coul.~gauge} &=&-ig \delta^{ij} U^{\dagger}_{\xt} \Big[\alpha^{i}_{\xt},\beta^{j}_{\xt}\Big] U_{\xt} +\mathcal{O}(\mathcal{A}_{(V)}^2)\;, \\
\left. B^{\eta}(\tau=0^{+},\xt) \right|_{Coul.~gauge}  &=&-ig \epsilon^{ij} U^{\dagger}_{\xt} \Big[\alpha^{i}_{\xt},\beta^{j}_{\xt}\Big] U_{\xt} +\mathcal{O}(\mathcal{A}_{(V)}^2);.
\end{eqnarray}       
Expressing $U^{\dagger}_{\xt} \Big[\alpha^{i}_{\xt},\beta^{j}_{\xt}\Big] U_{\xt} = \beta^{j,a}_{\xt} \Big(\partial_{i} U_{\xt}^{ab}\Big) t^{b}$ and performing the transformation to Fourier space, the field-strength bi-linears $\varepsilon(\tau,\xt)$ and $\nudot(\tau,\xt)$ can be compactly expressed  as
\begin{align}
&\varepsilon(\tau,\xt)=(-ig)^2 \int_{\pt,\pd} \int_{\kt,\kd}\Big(\delta^{ij}\delta^{kl} + \varepsilon^{ij} \varepsilon^{kl}\Big)  \frac{\beta^{j,a}_{\kt} (i\pt^{i}) U_{\pt}^{ab}(i\pd^{k})  U_{\pd}^{\dagger ba'}\beta^{l,a'}_{\kd} }{2} e^{i(\pt+\kt+\pd+\kd)\xt} \nonumber \\
&\qquad \qquad \times \left[ J_{0}( |\pt+\kt| \tau) J_{0}(|\pd+\kd| \tau) - \frac{(\pt+\kt) \cdot (\pd+\kd)}{|\pt+\kt| |\pd+\kd|} J_{1}(|\pt+\kt| \tau)J_{1}(|\pd+\kd| \tau)\right] \;, \nonumber \\
\\
&\nudot(\tau,\xt)=\frac{(ig)^2}{2} \int_{\pt,\pd} \int_{\kt,\kd}\Big(\delta^{ij}\varepsilon^{kl} + \varepsilon^{ij} \delta^{kl}\Big) \frac{\beta^{j,a}_{\kt} (i\pt^{i}) U_{\pt}^{ab}(i\pd^{k})  U_{\pd}^{\dagger ba'}\beta^{l,a'}_{\kd} }{2}  e^{i(\pt+\kt+\pd+\kd)\xt} \nonumber \\
&\qquad \qquad \times \left[ J_{0}( |\pt+\kt| \tau) J_{0}(|\pd+\kd| \tau) { -} \frac{(\pt+\kt) \cdot (\pd+\kd)}{|\pt+\kt| |\pd+\kd|} J_{1}(|\pt+\kt| \tau)J_{1}(|\pd+\kd| \tau)\right]   \;, \nonumber \\
\end{align}
where $\int_\kt$ stands for $\int \frac{\ud^2 \kt}{(2\pi)^2}$. Similarly, the two-point correlation functions of interest take the following form
\begin{align}
\label{eq:epsCorrMomentumSpace}
\varepsilon(\tau,\xt)\varepsilon(\tau,\yt)&=(ig)^4 \int_{\pt,\pd} \int_{\kt,\kd} \int_{\qt\qd} \int_{\lt,\ld} \Big(\delta^{ij}\delta^{kl} + \varepsilon^{ij} \varepsilon^{kl}\Big) \Big(\delta^{i'j'}\delta^{k'l'} + \varepsilon^{i'j'} \varepsilon^{k'l'}\Big)   \nonumber \\
&\times  \left\langle \left(  \frac{\beta^{j,a}_{\kt} (i\pt^{i}) U_{\pt}^{ab}(i\pd^{k})  U_{\pd}^{\dagger ba'}\beta^{l,a'}_{\kd} }{2} \right) 
 \left( \frac{\beta^{j',c}_{\lt} (i\qt^{i'}) U_{\qt}^{cd} (i\qd^{k'}) U_{\qd}^{\dagger dc'}\beta^{l',c'}_{\ld}}{2} \right) \right  \rangle  \nonumber  \\
&\times \left[ J_{0}( |\pt+\kt| \tau) J_{0}(|\pd+\kd| \tau) { -}  \frac{(\pt+\kt) \cdot (\pd+\kd)}{|\pt+\kt| |\pd+\kd|} J_{1}(|\pt+\kt| \tau)J_{1}(|\pd+\kd| \tau)\right] \nonumber \\
&\times \left[ J_{0}( |\qt+\lt| \tau) J_{0}(|\qd+\ld| \tau) { -}  \frac{(\qt+\lt) \cdot (\qd+\ld)}{|\qt+\lt| |\qd+\ld|} J_{1}(|\qt+\lt| \tau)J_{1}(|\qd+\ld| \tau)\right]      \nonumber \\
& \times e^{i(\pt+\kt+\pd+\kd)\xt} e^{i(\qt+\lt+\qd+\ld)\yt} \;.  \nonumber \\
\end{align}
\begin{align}
\label{eq:NuDotCorrMomentumSpace}
\nudot(\tau,\xt)\nudot(\tau,\yt)&=(ig)^4 \int_{\pt,\pd} \int_{\kt,\kd} \int_{\qt\qd} \int_{\lt,\ld} \Big(\delta^{ij}\varepsilon^{kl} + \varepsilon^{ij} \delta^{kl}\Big) \Big(\delta^{i'j'}\varepsilon^{k'l'} + \varepsilon^{i'j'} \delta^{k'l'}\Big)  \nonumber \\
&\times  \left\langle \left(  \frac{\beta^{j,a}_{\kt} (i\pt^{i}) U_{\pt}^{ab}(i\pd^{k})  U_{\pd}^{\dagger ba'}\beta^{l,a'}_{\kd} }{2} \right) 
 \left( \frac{\beta^{j',c}_{\lt} (i\qt^{i'}) U_{\qt}^{cd} (i\qd^{k'}) U_{\qd}^{\dagger dc'}\beta^{l',c'}_{\ld}}{2} \right) \right  \rangle  \nonumber  \\
&\times \left[ J_{0}( |\pt+\kt| \tau) J_{0}(|\pd+\kd| \tau) { -}  \frac{(\pt+\kt) \cdot (\pd+\kd)}{|\pt+\kt| |\pd+\kd|} J_{1}(|\pt+\kt| \tau)J_{1}(|\pd+\kd| \tau)\right] \nonumber \\
&\times \left[ J_{0}( |\qt+\lt| \tau) J_{0}(|\qd+\ld| \tau) { -}  \frac{(\qt+\lt) \cdot (\qd+\ld)}{|\qt+\lt| |\qd+\ld|} J_{1}(|\qt+\lt| \tau)J_{1}(|\qd+\ld| \tau)\right]   \nonumber \\
 & \times e^{i(\pt+\kt+\pd+\kd)\xt} e^{i(\qt+\lt+\qd+\ld)\yt} \;.  \nonumber \\
\end{align}
At early times $\tau\lesssim 1/\qs$ the products of Bessel functions are dominated by $J_0^2\approx 1$, corresponding to the limit discussed in Sec.~\ref{sec:correlators}. Beyond early times only the disconnected contribution has a delta function setting $\pd= -\pt, \ \kd = -\kt$ etc in such a way that the Bessel functions are arranged into combinations $J_0^2(x)+J_1^2(x)$ with the same argument $x$. Based on the approximate relation $J_0^2(x)+J_1^2(x) \approx 2/(\pi x)$ one then obtains the usual behavior of the energy density as $\varepsilon(\tau)\sim 1/\tau$. On the other hand, simplifications of this nature do not occur for the disconnected contributions, and the Bessel functions oscillate out of phase. Hence, we expect the correlation signal for the energy density and the divergence of the Chern-Simons current in coordinate space to vanish for $\tau \gg 1/\qs$. 

\subsection{Glasma graph two gluon correlation}
\label{sec:gg}

\begin{figure}[h!]
\begin{center}
\resizebox{5cm}{!}{
\includegraphics[width=6cm]{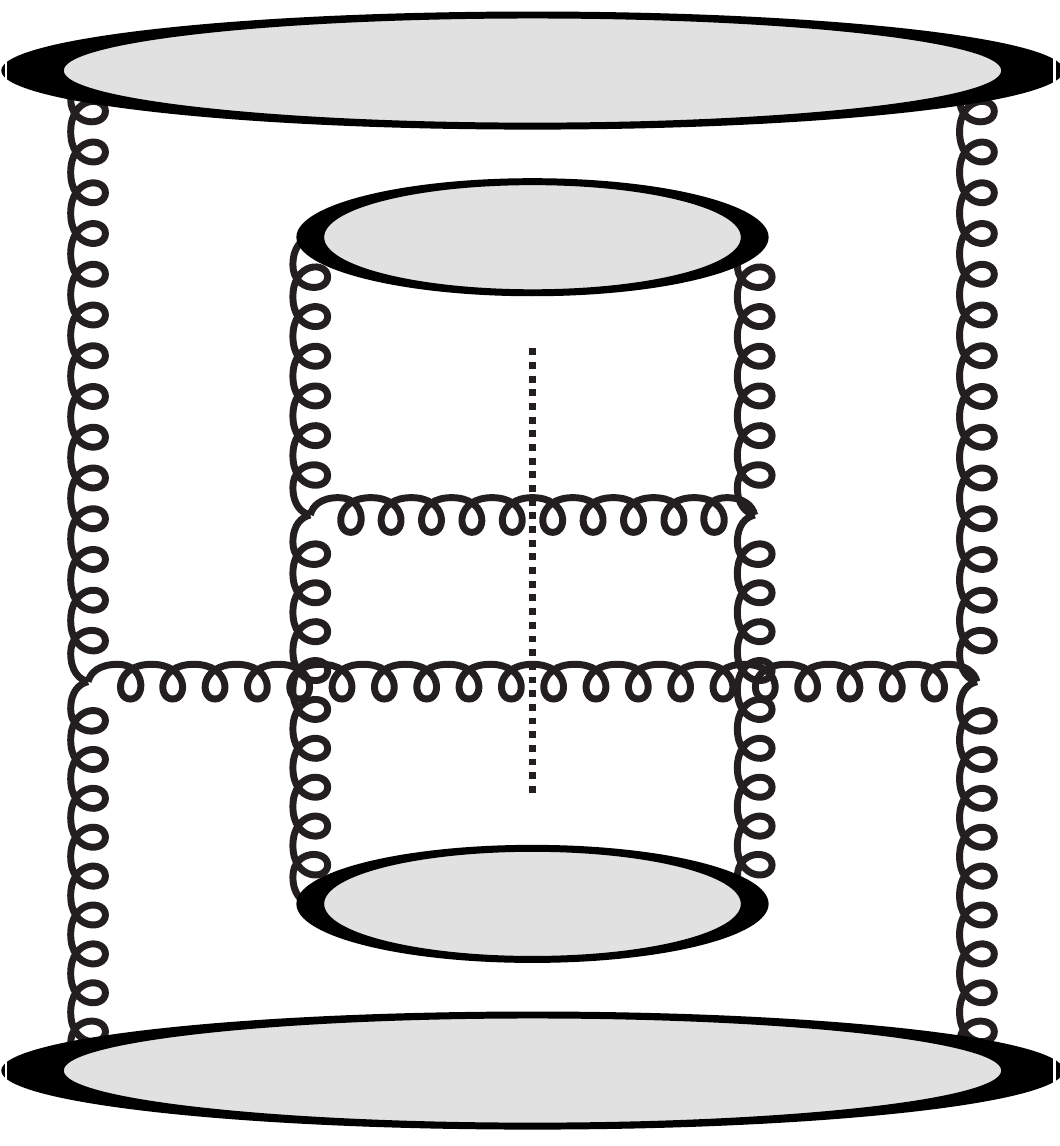}
\begin{tikzpicture}[overlay]
\node[anchor=east] at (-6.5,3) {\LARGE (a)};
\node[anchor=south east] at (-3.1,3.6) {$\yt$};
\node[anchor=south west] at (-3.1,3.6) {$\yd$};
\node[anchor=south east] at (-3.1,2.65) {$\xt$};
\node[anchor=south west] at (-3.1,2.65) {$\xd$};
\node[anchor=south east] at (-5.9,3.9) {$\pt$};
\node[anchor=north east] at (-5.9,2.0) {$\kt$};
\node[anchor=south east] at (-4.5,3.9) {$\qt$};
\node[anchor=north east] at (-4.5,2.0) {$\lt$};
\node[anchor=south west] at (-0.5,3.9) {$\pd$};
\node[anchor=north west] at (-0.5,2.0) {$\kd$};
\node[anchor=south west] at (-1.8,3.9) {$\qd$};
\node[anchor=north west] at (-1.8,2.0) {$\ld$};
\end{tikzpicture}
}
\end{center}

\begin{center}
\resizebox{5cm}{!}{
\includegraphics[width=6cm]{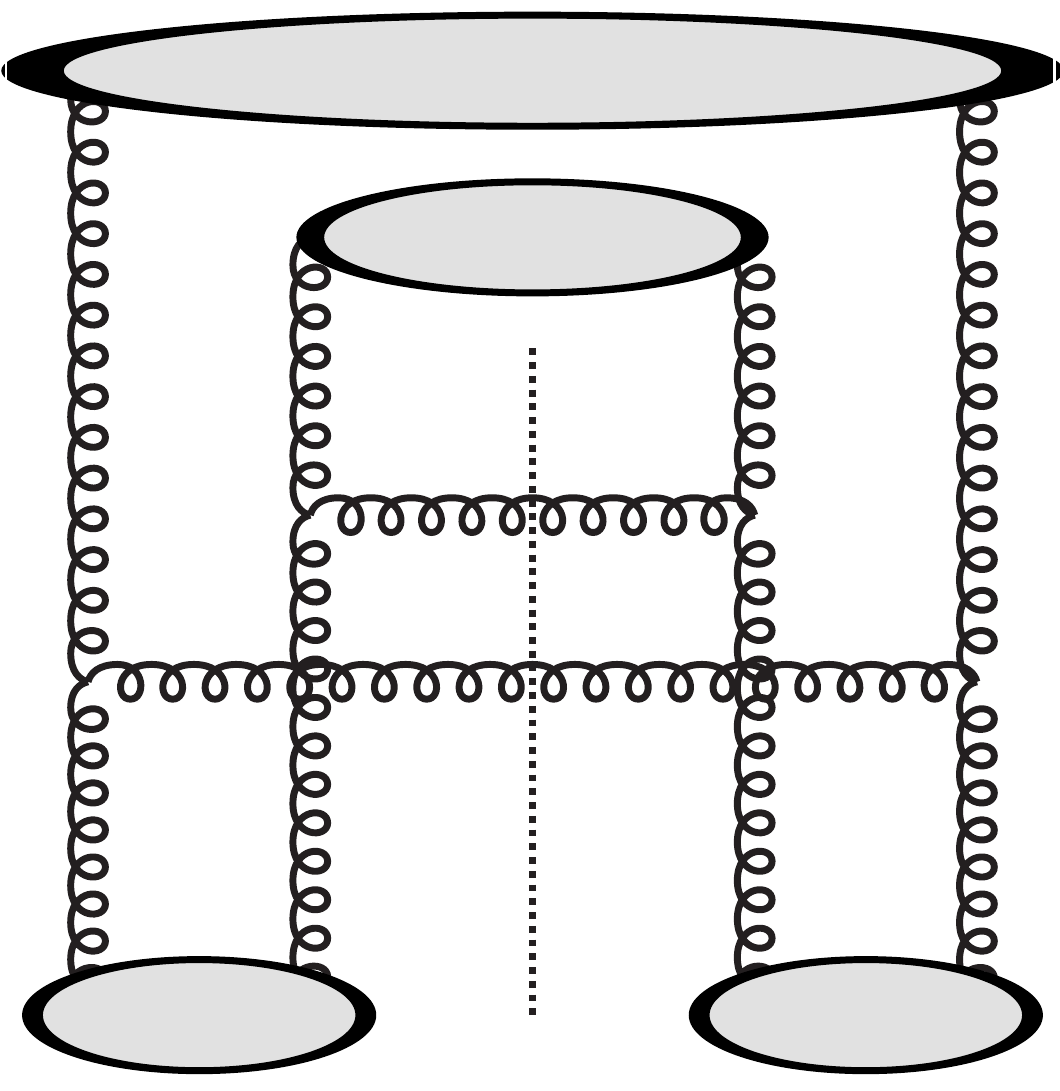}
\begin{tikzpicture}[overlay]
\node[anchor=east] at (-6.5,3) {\LARGE (b)};
\node[anchor=south east] at (-3.1,3.3) {$\yt$};
\node[anchor=south west] at (-3.1,3.3) {$\yd$};
\node[anchor=south east] at (-3.1,2.35) {$\xt$};
\node[anchor=south west] at (-3.1,2.35) {$\xd$};
\node[anchor=south east] at (-5.9,3.6) {$\pt$};
\node[anchor=north east] at (-5.9,1.7) {$\kt$};
\node[anchor=south east] at (-4.5,3.6) {$\qt$};
\node[anchor=north east] at (-4.5,1.7) {$\lt$};
\node[anchor=south west] at (-0.5,3.6) {$\pd$};
\node[anchor=north west] at (-0.5,1.7) {$\kd$};
\node[anchor=south west] at (-1.8,3.6) {$\qd$};
\node[anchor=north west] at (-1.8,1.7) {$\ld$};
\end{tikzpicture}
}
\end{center}

\begin{center}
\resizebox{5cm}{!}{
\includegraphics[width=6cm]{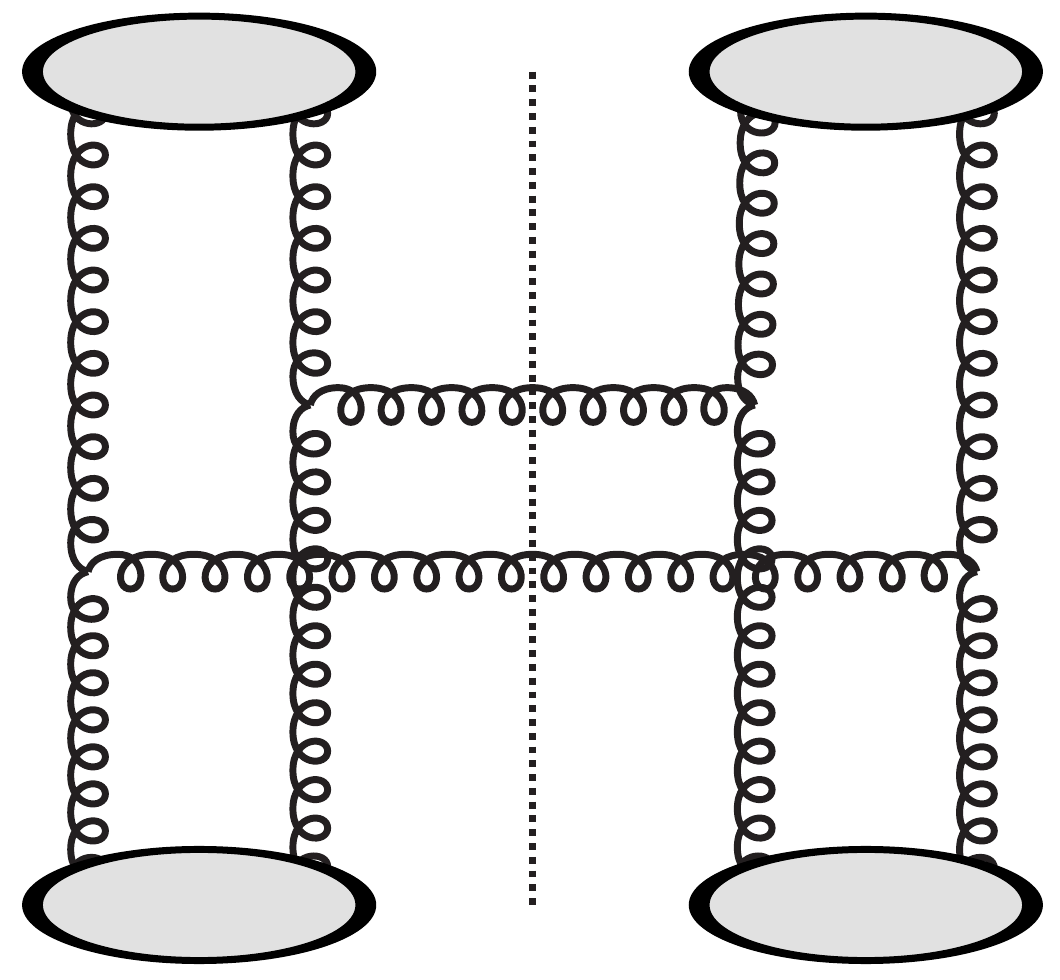}
\begin{tikzpicture}[overlay]
\node[anchor=east] at (-6.5,3) {\LARGE (c)};
\node[anchor=south east] at (-3.1,3.3) {$\yt$};
\node[anchor=south west] at (-3.1,3.3) {$\yd$};
\node[anchor=south east] at (-3.1,2.35) {$\xt$};
\node[anchor=south west] at (-3.1,2.35) {$\xd$};
\node[anchor=south east] at (-5.9,3.6) {$\pt$};
\node[anchor=north east] at (-5.9,1.7) {$\kt$};
\node[anchor=south east] at (-4.5,3.6) {$\qt$};
\node[anchor=north east] at (-4.5,1.7) {$\lt$};
\node[anchor=south west] at (-0.5,3.6) {$\pd$};
\node[anchor=north west] at (-0.5,1.7) {$\kd$};
\node[anchor=south west] at (-1.8,3.6) {$\qd$};
\node[anchor=north west] at (-1.8,1.7) {$\ld$};
\end{tikzpicture}
}\end{center}

\caption{\label{fig:Diagrams} Examples of (a) completely disconnected diagram and examples of (b) a disconnected-connected correlation (i.e. a ``rainbow diagram'' in the terminology of \cite{Gelis:2009wh}) and (c) a connected-connected correlation in the Glasma graph approximation.
For particle production, the coordinates in the amplitude $\xt,\yt$ are different from those in the conjugate amplitude, $\xd,\yd$, and are related by the momenta of the produced gluon. For the energy density and axial charge correlators, on the other hand, we integrate over momenta of the final state gluons, setting $\xt=\xd, \ \yd = \yt$.
}
\end{figure}

While the coordinate space correlation can be argued to vanish at $\tau \gg 1/\qs$, the situation is different for particle production, which is measured in momentum space. Here one integrates over the coordinates $\xt,\yt$ and the corresponding $\xd,\yd$ for the conjugate amplitude. This gives an additional momentum conservation delta function, which always sets $\pt+\kt = -(\pd+\kd)$ and  $\qt+\lt = -(\qd+\ld)$ also for the connected contributions in the expressions analogous to Eqs.~\nr{eq:epsCorrMomentumSpace}, \nr{eq:NuDotCorrMomentumSpace} (for an illustration of the momentum flow see Fig.~\ref{fig:Diagrams}). This leads to the ``glasma graph''~\cite{Dusling:2012iga} momentum space correlation structure. Even though this has not been the main focus of our paper, it is illustrative to derive this momentum space correlation signal here. This will clarify the relation of the calculation of Sec.~\ref{sec:correlators} to the earlier literature on these ``Glasma graph'' correlations~\cite{Dumitru:2008wn,Gelis:2009wh,Dusling:2009ni,Dumitru:2010iy,Dusling:2012iga,Dusling:2012wy,Dusling:2013qoz}.

In order to obtain single and double inclusive particle spectra at leading order accuracy in the LSZ formalism one usually considers the limit $\tau\to \infty$ and projects gauge fixed equal-time correlation functions onto plane wave modes $\xi^{\mu,(\lambda)}_{\kt}(\tau)$ according to
\begin{eqnarray}
\label{eq:lsz}
\frac{\ud N_{g}}{\ud y \ud^2\Pt} &=& \frac{1}{(2\pi)^2} \lim_{\tau \to \infty}  \sum_{\lambda,a} \left| \tau g^{\mu\nu} \left(\Big(\xi^{\Pt,(\lambda)}_{\mu}(\tau) \Big)^{*}~\overleftrightarrow{\partial_{\tau}}~A_{\nu}^{a}(\tau,\Pt) \right) \right|^2\;.
\end{eqnarray}
By use of the explicit form of the plane wave solutions in transverse Coulomb gauge~\cite{Makhlin:2000nx}
\begin{align}
\xi^{\kt,(1)}_{i}(\tau)&=\frac{\sqrt{\pi}}{2\ktt}  \varepsilon^{ij} k_{j} H_{0}^{(2)}(\ktt \tau)\;,  \qquad  \xi^{\kt,(2)}_{\eta}(\tau)=\frac{\sqrt{\pi}}{2\ktt} \ktt \tau H_{1}^{(2)}(\ktt \tau)\;.
\end{align}
with $\xi^{\kt,(1)}_{\eta}(\tau)=0$ and $\xi^{\kt,(2)}_{i}(\tau)=0$ and the orthonormality relations for Bessel type functions
\begin{eqnarray}
\Big(H^{(2)}_{0}(x) \Big)^{*}~\overleftrightarrow{\partial_{x}}~J_{0}(x) = -\frac{2 i }{\pi x}\;, \qquad  \Big(x H^{(2)}_{1}(x)  \Big)^{*}~\overleftrightarrow{\partial_{x}}~x J_{1}(x) = -\frac{2 i x}{\pi}\;,
\end{eqnarray}
the above expression evaluates to 
\begin{eqnarray}
\label{eq:sinc}
\frac{\ud N_{g}}{\ud y \ud^2\Pt} &=& \frac{1}{(2\pi)^2} \int_{\xt\xd} \frac{2}{\pi \Pt^2} \tr\Big( E^{\eta}(0^{+},\xt) E^{\eta}(0^{+},\xd) + B^{\eta}(0^{+},\xt) B^{\eta}(0^{+},\xd) \Big)_{Coul. gauge}~e^{-i\Pt(\xt-\xd)}  \nonumber \\
\end{eqnarray}
where in the dilute-dense regime the correlation functions in Coulomb gauge are given by
\begin{eqnarray}
&&\tr\Big( E^{\eta}(0^{+},\xt) E^{\eta}(0^{+},\xd) + B^{\eta}(0^{+},\xt) B^{\eta}(0^{+},\xd) \Big)_{Coul. gauge}=  \\
&&\qquad (-ig)^2  \int_{\pt,\pd} \int_{\kt,\kd} \Big(\delta^{ij}\delta^{kl} + \varepsilon^{ij} \varepsilon^{kl}\Big)    \left(  \frac{\beta^{j,a}_{\kt} (i\pt^{i}) U_{\pt}^{ab}(i\pd^{k})  U_{\pd}^{\dagger ba'}\beta^{l,a'}_{\kd} }{2} \right) e^{i(\pt+\kt)\xt} e^{i(\pd+\kd)\xd}\;. \nonumber
\end{eqnarray}

The single inclusive gluon spectrum is obtained by evaluating the expectation value of \nr{eq:sinc} directly in momentum space as
\begin{eqnarray}
&&(-ig)^2\left\langle\frac{\beta^{j,a}_{\kt} (i\pt^{i}) U_{\pt}^{ab}(i\pd^{k})  U_{\pd}^{\dagger ba'}\beta^{l,a'}_{\kd} }{2} \right \rangle=  \frac{g^2 N_c (N_c^2-1)}{2} ~  (2\pi)^2 \delta^{(2)}(\pt+\pd) (2\pi)^2 \delta^{(2)}(\kt+\kd) \nonumber \\
&&\qquad \qquad \qquad \times  \left[ \frac{\pt^{i}\pt^{k}}{\pt^2} D^{(1)}_{(U)}(|\pt|)\right] \times \left[\frac{1}{2} \delta^{jl} G^{(1)}_{(V)}(|\kt|)+\frac{1}{2} \Big(2\frac{\kt^{j}\kt^{l}}{\kt^2}-\delta^{jl} \Big) h^{(1)}_{\bot(V)}(|\kt|)\right]
\end{eqnarray}
where we defined the Dipole gluon distribution\footnote{Note that the normalization of the dipole distribution is chosen such that in the dilute limit of the McLerran-Venugopalan model, the gluon distributions are all equal $D^{(1)}(|\kt|)=G^{(1)}(|\kt|)=h^{(1)}(|\kt|)=g^2 \frac{\mu^2(\kt)}{\kt^2}$.}
\begin{eqnarray}
D^{(1)}_{(U)}(|\pt|)=\frac{\pt^2}{S_{\bot} \nc}\int_{\xt,\xd} \frac{1}{N_c^2-1} \tr[U_{\xt}^{adj} U^{\dagger,adj}_{\xd}] e^{i\pt(\xt-\xd)} \;.
\end{eqnarray}
Note that the dipole distribution is explicitly proportional to the momentum. Thus in a decomposition into polarization states similarly as for the Weisz\"{a}cker-Williams distribution, the unpolarized and polarized distributions are equal and there is only one scalar distribution $D^{(1)}$, with
\begin{eqnarray}
D^{ik}_{(U)}(\pt)=\frac{\pt^{i}\pt^{k}}{\pt^2} D^{(1)}_{(U)}(|\pt|) \; .
\end{eqnarray}
Using these expressions we obtain the following result for the single inclusive distribution
\begin{eqnarray}
\frac{\ud N_{g}}{\ud^2\Pt \ud y} = \frac{g^2 \nc (\nc^2-1)}{(2\pi)^2} \frac{S_{\bot}}{\pi \Pt^2}  
\int \frac{\ud^2\kt}{(2\pi)^2} D^{(1)}_{(U)}(\kt) G^{(1)}_{(V)}(\Pt-\kt)\;.
\end{eqnarray}

We are now in a situation  to repeat the calculation of double inclusive gluon production in \cite{Dusling:2009ni} in our notations. Since the Glasma graph contribution to the double inclusive spectrum is simply given by the square of the single inclusive spectrum, we  obtain
\begin{eqnarray}
\label{eq:DoubleInclusive}
\left\langle \frac{\ud N_{g}}{\ud y_P \ud^2\Pt \ud y_Q \ud^2\Qt} \right  \rangle&=&\frac{(-ig)^4 }{(2\pi)^4} \int_{\xt\xd} \int_{\yt\yd} \frac{2 e^{-i\Pt(\xt-\xd)} }{\pi \Pt^2} \frac{2 e^{-i\Qt(\yt-\yd)} }{\pi \Qt^2}  \int_{\pt,\pd} \int_{\kt,\kd}  \int_{\qt\qd} \int_{\lt,\ld}   \\
&&   \times \Big(\delta^{ij}\delta^{kl} + \varepsilon^{ij} \varepsilon^{kl}\Big)    \Big(\delta^{i'j'}\delta^{k'l'} + \varepsilon^{i'j'} \varepsilon^{k'l'}\Big)  e^{i(\pt+\kt)\xt} e^{i(\pd+\kd)\xd} e^{i(\qt+\lt)\yt} e^{i(\qd+\ld)\yd}   \nonumber \\
&& \times  \left\langle \left(  \frac{\beta^{j,a}_{\kt} (i\pt^{i}) U_{\pt}^{ab}(i\pd^{k})  U_{\pd}^{\dagger ba'}\beta^{l,a'}_{\kd} }{2} \right) 
 \left( \frac{\beta^{j',c}_{\lt} (i\qt^{i'}) U_{\qt}^{cd} (i\qd^{k'}) U_{\qd}^{\dagger dc'}\beta^{l',c'}_{\ld}}{2} \right) \right  \rangle\;. \nonumber
\end{eqnarray}
where we inserted the explicit expressions for the chromo-electric and chromo-magnetic fields in order to make the similarities and differences with Eq.~(\ref{eq:NuDotCorrMomentumSpace}) most apparent.

One immediately observes that both Eq.~(\ref{eq:NuDotCorrMomentumSpace}) and Eq.~(\ref{eq:DoubleInclusive}) involve the same correlation function of the gluon fields in momentum-space, allowing for the same interpretation in terms of a diagrammatic analysis. Specifically, the various different contractions in the projectile and target fields can be associated with the usual Glasma graphs as illustrated in Fig.~(\ref{fig:Diagrams}). Even though the diagrammatics is essentially the same for double inclusive production and two-point correlation functions of local operators, there are of course some crucial differences in the calculation. Besides the appearance of a different operator structure in the middle of Eq.~(\ref{eq:DoubleInclusive}), another key difference is that for the local operator correlation function $\langle \nudot(\xt)\nudot(\yt) \rangle$ (and similarly for $\langle \varepsilon(\xt)\varepsilon(\yt) \rangle$) all expressions are to be evaluated at the same coordinates $\xt=\xd$ and $\yt=\yd$ in the amplitude and complex conjugate amplitude. 
Moreover, for the double inclusive spectrum the relevant correlation function of Wilson lines is given by 
\begin{equation}
Q^{iki'k'}_{jlj'l'}(\pt,\pd;\kt,\kd)=(-ig)^4  \left\langle \left(  \frac{\beta^{j,a}_{\kt} (i\pt^{i}) U_{\pt}^{ab}(i\pd^{k})  U_{\pd}^{\dagger ba'}\beta^{l,a'}_{\kd} }{2} \right) 
 \left( \frac{\beta^{j',c}_{\lt} (i\qt^{i'}) U_{\qt}^{cd} (i\qd^{k'}) U_{\qd}^{\dagger dc'}\beta^{l',c'}_{\ld}}{2} \right) \right  \rangle\;,
\end{equation}
with the crucial difference that the contractions on the dense side now involve adjoint Wilson lines $U^{ab}$, instead of the Weisz\"{a}cker-Williams field $\alpha^{i}_{(U)}$ as discussed in Sec.~\ref{sec:correlators}. Generally speaking the expectation value of the four point correlation function of adjoint Wilson lines
\begin{eqnarray}\label{eq:4ptadj}
\left\langle U_{\pt}^{ab} U_{\pd}^{\dagger ba'} U_{\qt}^{cd} U_{\qd}^{\dagger dc'}  \right\rangle 
\end{eqnarray}
can be decomposed into a complete set of color singlet structures~\cite{Fukushima:2007dy}. Evaluating the full color structure is however quite challenging, and following~\cite{Dumitru:2008wn,Gelis:2009wh,Dusling:2009ni} one usually resorts to an approximation of the full color structure in terms of the leading components in a dilute expansion. Specifically one expands the adjoint Wilson lines in Eq.~\nr{eq:4ptadj} to lowest order in the target fields
\begin{eqnarray}\label{eq:dip4pt}
\left\langle U_{\pt}^{ab} U_{\pd}^{\dagger ba'} U_{\qt}^{cd} U_{\qd}^{\dagger dc'}  \right\rangle   \simeq  
\left\langle\Big(\delta^{ab} \delta(\pt) + i g \mathcal{A}^{(U)}_{e}(\pt)  T_{e}^{~ab}  + \cdots\Big)~
\Big(\delta^{ba'}\delta(\pd) - i g \mathcal{A}^{(U)}_{e'}(\pd) T_{e'}^{~ba'} + \cdots\Big) \right. \nonumber \\
\left. \Big(\delta^{cd} \delta(\qt) + i g \mathcal{A}^{(U)}_{f}(\qt)  T_{f}^{~cd}  + \cdots\Big)~
\Big(\delta^{dc'}\delta(\qd) - i g \mathcal{A}^{(U)}_{f'}(\qd) T_{f'}^{~dc'} + \cdots\Big) \right \rangle \nonumber \\
\end{eqnarray}
and performs a Gaussian averaging in terms of the fields $\mathcal{A}^{(U)}$ according to
\begin{eqnarray}
g^2 \left \langle \mathcal{A}^{(U)}_{e}(\pt)  \mathcal{A}^{(U)}_{e'}(\pd) \right \rangle = (2\pi)^2 \delta(\pt+\pd) \frac{D^{(1)}_{(U)}(\pt)}{\pt^2} \delta_{ee'}\;.
\end{eqnarray}
Similar to the discussion in Sec.~\ref{sec:correlators}, the correlation function can then be evaluated in terms of the disconnected-disconnected ($DD$), disconnected-connected ($DC$),  connected-disconnected ($CD$) and connected-connected ($CC$) contributions
\begin{eqnarray}
Q^{iki'k'}_{jlj'l'}(\pt,\pd;\kt,\kd)=DD + DC+CD+CC\;.
\end{eqnarray}
Expressing the adjoint generators explicitly as $T_{e}^{~ab}=if^{abe}$ the color factors are exactly the same ones as in Eq.~\nr{eq:ww4pt}, and it is then straightforward to obtain
\begin{align}
\label{eq:ggcontractions}
DD &=\frac{g^4 \nc^2 (\nc^2-1)^2}{4} 
(2\pi)^{2} \delta^{(2)}(\pt+\pd) D_{(U)}^{ik}(\pt) ~ (2\pi)^{2} \delta^{(2)}(\qt+\qd)D_{(U)}^{i'k'}(\qt) \nonumber \\
 & \qquad \qquad \qquad \quad~\times~(2\pi)^{2} \delta^{(2)}(\kt+\kd) W_{(V)}^{jl}(\kt) ~(2\pi)^2 \delta^{(2)}(\lt+\ld)  W_{(V)}^{j'k'}(\lt)\; \nonumber \\
\end{align}
\begin{align}
\label{eq:ggcontractions1}
DC&=\frac{g^4 \nc^2 (\nc^2-1)}{4}
(2\pi)^{2} \delta^{2}(\pt+\pd) D_{(U)}^{ik}(\pt) ~(2\pi)^{2} \delta^{2}(\qt+\qd) D_{(U)}^{i'k'}(\qt) \nonumber \\ 
& \qquad \qquad  \times \bigg[ (2\pi)^{2}\delta^{(2)}(\kt+\lt) W_{(V)}^{jj'}(\kt)~(2\pi)^{2}\delta^{(2)}(\kd+\ld) W_{(V)}^{ll'}(\kd) 
\nonumber \\
&
\qquad \qquad \qquad \qquad \qquad \qquad
+ (2\pi)^{2}\delta^{(2)}(\kt+\ld) W_{(V)}^{jl'}(\kt)~(2\pi)^{2}\delta^{(2)}(\lt+\kd) W_{(V)}^{lj'}(\kd) \bigg]  \nonumber \\
\end{align}
\begin{align}
\label{eq:ggcontractions2}
CD&=\frac{g^4 \nc^2 (\nc^2-1)}{4}
(2\pi)^{2}\delta^{(2)}(\kt+\kd) W_{(V)}^{jl}(\kt) ~(2\pi)^{2}\delta^{(2)}(\lt+\ld)  W_{(V)}^{j'l'}(\lt)  \nonumber \\ 
&\qquad \qquad \times \bigg[ (2\pi)^{2}\delta^{(2)}(\pt+\qt) D_{(U)}^{ii'}(\pt)~(2\pi)^{2}\delta^{(2)}(\pd+\qd) D_{(U)}^{kk'}(\pd
) 
\nonumber \\
&
\qquad \qquad \qquad \qquad \qquad \qquad
+ (2\pi)^{2}\delta^{(2)}(\pt+\qd) D_{(U)}^{ik'}(\pt) (2\pi)^{2}\delta^{(2)}(\qt+\pd) D_{(U)}^{ki'}(\pd) \bigg] \nonumber \\
\end{align}
\begin{align}
\label{eq:ggcontractions3}
CC&=  \frac{g^4 \nc^2 (\nc^2-1)}{4}  
\bigg\{ (2\pi)^{2}\delta^{(2)}(\pt+\qt)  D_{(U)}^{ii'}(\pt)~(2\pi)^{2}\delta^{(2)}(\pd+\qd) D_{(U)}^{kk'}(\pd)  
\nonumber \\
& 
\qquad \qquad 
\times \Big[ (2\pi)^{2}\delta^{(2)}(\kt+\lt) W_{(V)}^{jj'}(\kt) (2\pi)^{2}\delta^{(2)}(\kd+\ld) W_{(V)}^{ll'}(\kd) 
\nonumber \\
&
\qquad \qquad \qquad \qquad \qquad 
+\frac{1}{2} (2\pi)^{2}\delta^{(2)}(\kt+\ld) W_{(V)}^{jl'}(\kt) (2\pi)^{2}\delta^{(2)}(\lt+\kd) W_{(V)}^{lj'}(\kd) \Big]  \nonumber \\
&   + (2\pi)^{2}\delta^{(2)}(\pt+\qd)  D_{(U)}^{ik'}(\pt) (2\pi)^{2}\delta^{(2)}(\qt+\pd) D_{(U)}^{ki'}(\pd)  \nonumber \\
& 
\qquad \qquad 
\times \Big[ \frac{1}{2} (2\pi)^{2} \delta^{(2)}(\kt+\lt) W_{(V)}^{jj'}(\kt) (2\pi)^{2} \delta^{(2)}(\kd+\ld) W_{(V)}^{ll'}(\kd) 
\nonumber \\
&
\qquad \qquad \qquad \qquad \qquad 
+  (2\pi)^{2}\delta^{(2)}(\kt+\ld) W^{jl'}(\kt) (2\pi)^{2}\delta^{(2)}(\lt+\kd) W^{lj'}(\kd) \Big] \bigg\} . \nonumber \\
\end{align}
where we dropped the contributions of all terms proportional to delta functions of a single momentum, i. e. $\delta(\pt),\delta(\pd),\delta(\qt),\delta(\qd)$, as these do not contribute to particle production. We note that the approximation of the adjoint four point function in Eqns.~\nr{eq:ggcontractions}, \nr{eq:ggcontractions1}, \nr{eq:ggcontractions2}, \nr{eq:ggcontractions3} is equivalent to the approximation used for the four-point function of the Weisz{\"a}cker-Williams field in Eq.~\nr{eq:ww4pt} to leading order in the dilute limit. However, they correspond to different selective resummations of higher order terms away from the dilute limit. Ultimately this difference originates in the approximations used for the higher point functions of Wilson lines in  Eqs.~\nr{eq:DoublePartonFactorization} and~\nr{eq:dip4pt}.
We also stress that one cannot perform a  naive decomposition of the four-point function of adjoint Wilson lines~\nr{eq:4ptadj} into pairwise contractions of nonsinglet 2-point functions. Such a procedure would, for example,  not reproduce the correct $\nc$~counting in the high momentum dilute limit, which we can check using the dilute approximation. Thus the ``glasma graph'' appproximation must be used with care, since it only really works in the dilute limit. In particular, we have not been able to find  a $k_T$-factorized expression for the 2-particle correlation function in the dilute-dense pA-case although one, involving the dipole distribution, does exist for the single gluon cross section.

Evaluating the individual terms, we obtain the following result for the contributions to the double-inclusive spectrum
\begin{multline}
\label{eq:DoubleInclusivePol}
\left\langle \frac{\ud N_{g}}{\ud^2\Pt \ud y_P \ud^2\Qt \ud y_Q} \right  \rangle - \left\langle \frac{\ud N_{g}}{\ud^2\Pt \ud y_P}  \right  \rangle  \left\langle \frac{\ud N_{g}}{\ud^2\Qt \ud y_Q} \right  \rangle
\\
=\frac{g^4 \nc^2 (\nc^2-1)}{ (2\pi)^4} 
S_{\bot} \frac{1}{\pi \Pt^2} \frac{1}{\pi \Qt^2} \left( DC+CD+CC_{S} +CC_{A}\right)  
\end{multline}
where the "disconnected-connected" ($DC$), "connected-disconnected" ($CD$) and (symmetry/asymmetric) "connected-connected" ($CC_{S}/CC_{A}$) contributions are given by
\begin{align}
\label{eq:DoubleInclusivePol1}
DC&= \frac{1}{2} \int \frac{\ud^2\kt}{(2\pi)^2} \left[ D^{(1)}_{(U)}(\Pt-\kt) D^{(1)}_{(U)}(\Qt+\kt) + D^{(1)}_{(U)}(\Pt-\kt) D^{(1)}_{(U)}(\Qt-\kt)  \right] \\
& \qquad \qquad\qquad \qquad \times \left[  \left(G^{(1)}_{(V)}(\kt) \right)^2 
+ \left(h^{(1)}_{\bot(V)}(\kt)\right)^2 \right] \nonumber  \\
\nonumber \\
\label{eq:DoubleInclusivePol2}
CD &= \int \frac{\ud^2\kt}{(2\pi)^2} \left(D^{(1)}_{(U)}(\kt)\right)^2  \left[ G^{(1)}_{(V)}(\Pt-\kt) G^{(1)}_{(V)}(\Qt+\kt) + G^{(1)}_{(V)}(\Pt-\kt) G^{(1)}_{(V)}(\Qt-\kt)  \right]   \\
\nonumber \\
\label{eq:DoubleInclusivePol3}
CC_{S} &= \frac{1}{2}  \left( (2\pi)^2 \delta^{(2)}(\Pt+\Qt) + (2\pi)^2\delta^{(2)}(\Pt-\Qt) \right) \int \frac{\ud^2\kt}{(2\pi)^2}  \frac{\ud^2\kd}{(2\pi)^2}  D^{(1)}_{(U)}(\kt) D^{(1)}_{(U)}(\kd)  \nonumber \\
& \times \left[G^{(1)}_{(V)}(\Pt-\kt) G^{(1)}_{(V)}(\Pt+\kd) + h^{(1)}_{\bot(V)}(\Pt-\kt) h^{(1)}_{\bot(V)}(\Pt+\kd) \cos\big(2 (\theta_{\kt,\Pt-\kt} -\theta_{\kd,\Pt+\kd})\big) \right] \nonumber  \\
 \\
\label{eq:DoubleInclusivePol4}
CC_{A} &= \frac{1}{4} \int \frac{\ud^2\kt}{(2\pi)^2} \left[ D^{(1)}_{(U)}(\kt) D^{(1)}_{(U)}(\Pt-\Qt-\kt) h^{(1)}_{\bot(V)}(\Pt-\kt) h^{(1)}_{\bot(V)}(\Qt+\kt) \cos\big(2 \theta_{\Pt-\kt,\Qt+\kt}\big)   \right. \nonumber \\
& \qquad \qquad \quad + \left. D^{(1)}_{(U)}(\kt) D^{(1)}_{(U)}(\Pt-\Qt-\kt) ~ G^{(1)}_{(V)}(\Pt-\kt) G^{(1)}_{(V)}(\Qt+\kt) ~\cos\big(2 \theta_{\kt,\Pt-\Qt-\kt}\big)  \right. \nonumber \\
& \qquad \qquad \quad + \left. D^{(1)}_{(U)}(\kt) D^{(1)}_{(U)}(\Pt+\Qt-\kt) h^{(1)}_{\bot(V)}(\Pt-\kt) h^{(1)}_{\bot(V)}(\Qt-\kt) \cos\big(2 \theta_{\Pt-\kt,\Qt-\kt}\big) \right. \nonumber \\
&\qquad \qquad \quad + \left. D^{(1)}_{(U)}(\kt) D^{(1)}_{(U)}(\Pt+\Qt-\kt) ~G^{(1)}_{(V)}(\Pt-\kt) G^{(1)}_{(V)}(\Qt-\kt) ~\cos\big(2 \theta_{\kt,\Pt+\Qt-\kt}\big)  \right] .  \nonumber \\
\end{align}
One interesting feature of Eqs.~\nr{eq:DoubleInclusivePol},\nr{eq:DoubleInclusivePol1},  \nr{eq:DoubleInclusivePol2}, \nr{eq:DoubleInclusivePol3} and \nr{eq:DoubleInclusivePol4}  -- which is also visible in Eq.~\nr{eq:epsepscorr} -- is the polarization structure on the proton side, which we maintained in full generality. One sees that for the ``disconnected'' contribution on the proton side (the $CD$-term) only the unpolarized gluon distribution appears, whereas on the ``connected'' side  ($DC$ and $CC$-terms) one is sensitive to the sum of the squares of the unpolarized and linearly polarized distributions. This is a subtle effect of a full treatment  of a nontrivial linear polarization structure  on the two-gluon correlations in momentum space. \\

Our result in Eqs.~\nr{eq:DoubleInclusivePol}, \nr{eq:DoubleInclusivePol1}, \nr{eq:DoubleInclusivePol2} and \nr{eq:DoubleInclusivePol3} should be compared with the Glasma graph approximation originally derived in \cite{Dumitru:2008wn} (see also \cite{Gelis:2009wh},  Eq. (3.17) of \cite{Dusling:2009ni} and Eq.~(3) of \cite{Dumitru:2010iy} and Eqs. (1) and (2) of  \cite{Dusling:2012iga,Dusling:2012wy,Dusling:2013qoz} which correct typos in earlier references, see footnote [23] of \cite{Dusling:2012iga})). These assumed a maximal linear polarization $G^{(1)}(\kt) = h^{(1)}_{\perp}(\kt)$ for both projectiles, with
\begin{eqnarray}
G^{(1)}_{(U/V)}(\kt)=h^{(1)}_{\bot,(U/V)}(\kt)= D^{(1)}_{(U/V)}(\kt)= g^2~\frac{\mu_{(U/V)}^{2}(\kt)}{\kt^2}\;.
\end{eqnarray}
By expressing our results in this limit of full linear polarization in  terms of the gluon distribution normalized as in \cite{Dusling:2012iga,Dusling:2012wy,Dusling:2013qoz}
\begin{equation}
 G^{(1)}_{(U/V)}(\kt) = h^{(1)}_{\perp}(\kt)=  \frac{\mathbf{\Phi}_{U/V}(\kt)}{\pi (\nc^2-1)}
 \end{equation}
we obtain the well known $\kt_{\bot}$-factorization result for the single inclusive spectrum in the form quoted in \cite{Dusling:2012iga,Dusling:2012wy,Dusling:2013qoz}
\begin{eqnarray}
\frac{\ud N_{g}}{\ud^2\Pt \ud y} = \frac{\as \nc }{ \pi^4 (\nc^2-1)} \frac{S_{\bot}}{\Pt^2}  \int \frac{\ud^2\kt}{(2\pi)^2} \Phi_{U}(\kt) \Phi_{V}(\Pt-\kt).
\end{eqnarray}
Similarly -- omitting the connected-connected contributions for compactness -- our result for the double inclusive spectrum in the maximally polarized limit  reduces to the form used in Ref.~\cite{Dusling:2012iga,Dusling:2012wy,Dusling:2013qoz}:
\begin{eqnarray}
&&\left\langle \frac{\ud N_{g}}{\ud^2\Pt \ud y_P \ud^2\Qt \ud y_Q} \right  \rangle - \left\langle \frac{\ud N_{g}}{\ud^2\Pt\ud y _P}  \right  \rangle  \left\langle \frac{\ud N_{g}}{\ud^2\Qt\ud y_Q} \right  \rangle
=
\frac{\as^2 \nc^2}{ 4 \pi^{10} (\nc^2-1)^3}  \frac{S_{\bot}}{ \Pt^2 \Qt^2}
 \nonumber \\
&&
\times \int \ud^2\kt
\Big(~\left[ \Phi_{U}(\Pt-\kt) \Phi_{U}(\Qt+\kt) + \Phi_{U}(\Pt-\kt) \Phi_{U}(\Qt-\kt)  \right]~\Phi_{V}(\kt) \Phi_{V}(-\kt)  \nonumber \\
&&~+~\Phi_{U}(\kt) \Phi_{U}(-\kt) ~\left[ \Phi_{V}(\Pt-\kt) \Phi_{V}(\Qt+\kt) + \Phi_{V}(\Pt-\kt) \Phi_{V}(\Qt-\kt)  \right]\Big).
\end{eqnarray}

\section{Discussion}
\label{sec:phenomenology}
\begin{figure}[t!]
\begin{center}
\begin{minipage}{0.49\textwidth}
\includegraphics[width=\textwidth]{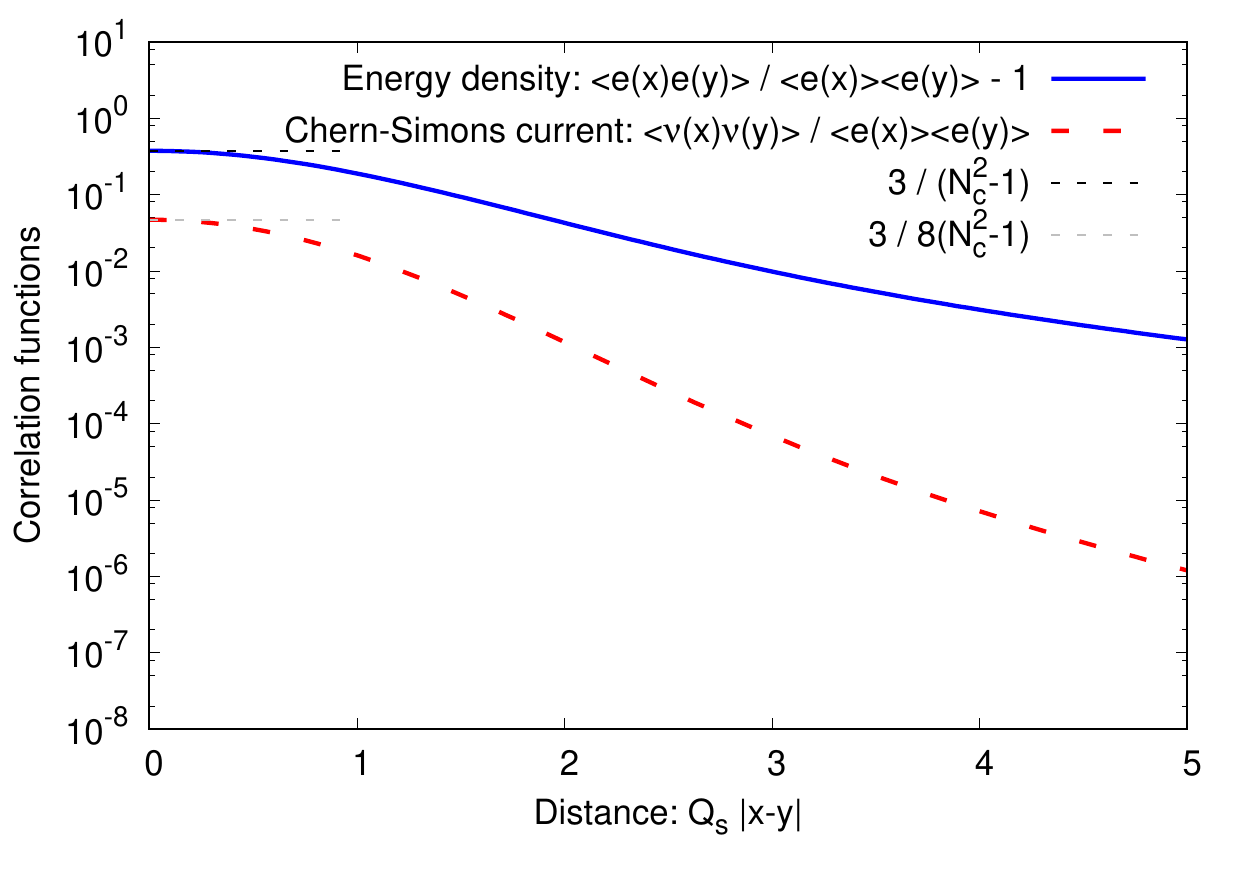}
\end{minipage}
\begin{minipage}{0.49\textwidth}
\includegraphics[width=\textwidth]{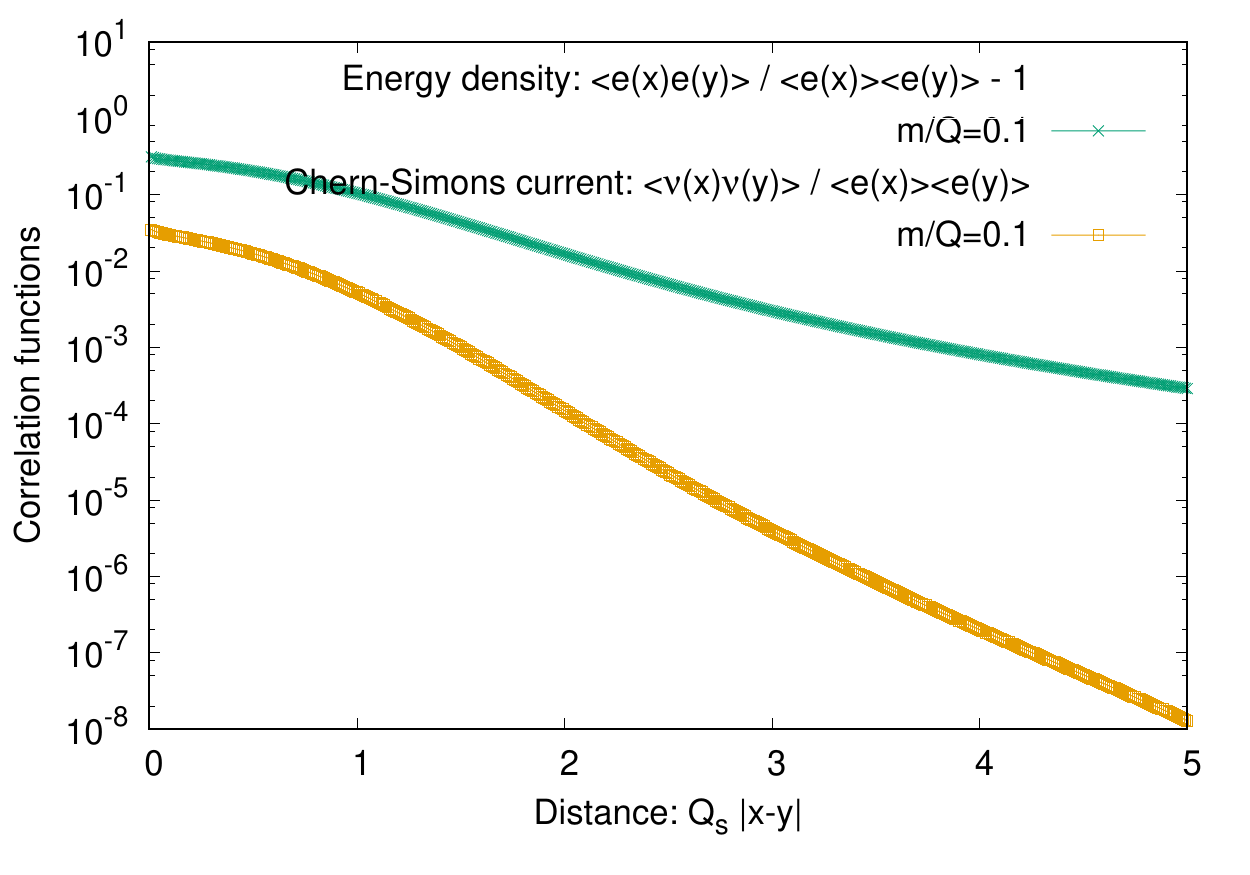}
\end{minipage}
\caption{\label{fig:Correlators} Comparison of the correlation functions of the energy density $\langle\varepsilon(\xt)\varepsilon(\yt)\rangle$  and the divergence of the Chern-Simons current $\langle \nudot(\xt) \nudot(\yt) \rangle$ in the GBW model (left) and MV model (right).}
\end{center}
\end{figure}

We now return to the central objective of this paper -- to characterize axial charge production in the Glasma. Based on our calculation in Sec.~\ref{sec:correlatorcalculation}, we find that the expectation value of the divergence of the Chern-Simons current $\langle \nudot(\xt) \rangle=0$ vanishes identically, such that on average no imbalance axial charge imbalance is created. However, the variance $\langle \nudot(\xt) \nudot(\yt) \rangle$ is finite, such that local fluctuations of the axial charge density should be expected on an event-by-event basis. Specifically, in the GBW saturation model, we obtain the following result for the correlation functions 
\begin{eqnarray}
\label{eq:EnergyFluctuationResult}
\frac{\langle \varepsilon(\xt) \varepsilon(\yt) \rangle}{\langle\varepsilon(\xt) \rangle \langle \varepsilon(\yt)\rangle} -1 &=& \frac{3}{\nc^2-1} \left[ \frac{1}{3} \left(\frac{ 1-e^{-\frac{\nc}{4\cf} \qs^2 |\xt-\yt|^2}}{\frac{\nc }{4\cf} \qs^2 |\xt-\yt|^2 }\right)^4 + \frac{2}{3} \left(\frac{ 1-e^{-\frac{\nc}{4\cf} \qs^2 |\xt-\yt|^2}}{\frac{\nc }{4\cf} \qs^2 |\xt-\yt|^2 }\right)^2 \right]\;, \nonumber \\
\end{eqnarray}
\begin{eqnarray}
\frac{\langle \nudot(\xt) \nudot(\yt) \rangle}{\langle\varepsilon(\xt) \rangle \langle \varepsilon(\yt)\rangle} &=& \frac{3}{8 (\nc^2-1)} \left(\frac{ 1-e^{-\frac{\nc}{4\cf} \qs^2 |\xt-\yt|^2}}{\frac{\nc }{4\cf} \qs^2 |\xt-\yt|^2 }\right)^4\;, 
\end{eqnarray}
which is depicted in the left panel of Fig.~\ref{fig:Correlators}. We note that except for the $1/(\nc^2-1)$ suppression factor characteristic for fluctuations, there is no parametric suppression of $\langle \nudot(\xt) \nudot(\yt) \rangle$ compared to the energy density $\langle\varepsilon(\xt) \rangle$, indicating that \emph{locally} Glasma flux tubes can induce a significant imbalance of the axial charge density. However, it is also evident from Fig.~\ref{fig:Correlators} that the correlation length of these Glasma flux tubes in the transverse plane is microscopically small $\sim 1/\qs$ -- such that a large number of uncorrelated domains should be expected in a realistic event. Besides the analytic results obtained in the GBW saturation model, the right panel of Fig.~\ref{fig:Correlators} shows the same quantities calculated in the MV model (see Sec.~\ref{sec:dipmodels} for details). 

Based on the above results for the source for axial charge production, we can also estimate local fluctuations of the axial charge density. Using our approximate treatment in Eq.~(\ref{eq:AxialChargeProduction}) we find that for times $\tau \lesssim 1/\qs$ the local fluctuations can be estimated as
\begin{eqnarray}
\label{eq:KeyResult}
\left\langle \frac{\ud N_5}{\ud^2\xt \ud\eta} \frac{\ud N_5}{\ud^2\yt \ud\eta} \right \rangle  \approx \frac{3\as^2 N_f^2}{8 \pi^2 (\nc^2-1)}~\langle\varepsilon(\xt) \rangle \langle \varepsilon(\yt)\rangle~ \tau^4~\left(\frac{ 1-e^{-\frac{\nc}{4\cf} \qs^2 |\xt-\yt|^2}}{\frac{\nc }{4\cf} \qs^2 |\xt-\yt|^2 }\right)^4\;.
\end{eqnarray}
whereas fluctuations of the global amount of axial charge are suppressed by the overall number of Glasma flux tubes $1/\qs^2 S_{\bot}$ and approximately given by
\begin{eqnarray}
\left\langle \int_{\xt} \frac{\ud N_5}{\ud^2\xt \ud\eta} \int_{\yt} \frac{\ud N_5}{\ud^2\yt \ud\eta} \right \rangle \approx  \kappa~\frac{\as^2 N_f^2}{2\pi^2 \nc^2}~\frac{\varepsilon^2 \tau^4 S_{\bot}^2}{\qs^2 S_{\bot}}\;,
\end{eqnarray}
where $\kappa=\pi (44 \ln(2) -27 \ln(3)) \approx 2.6262$ and $S_\bot$ denotes the transverse size of the overlap area. 

\begin{figure}[t!]
\begin{center}
\includegraphics[width=0.6\textwidth]{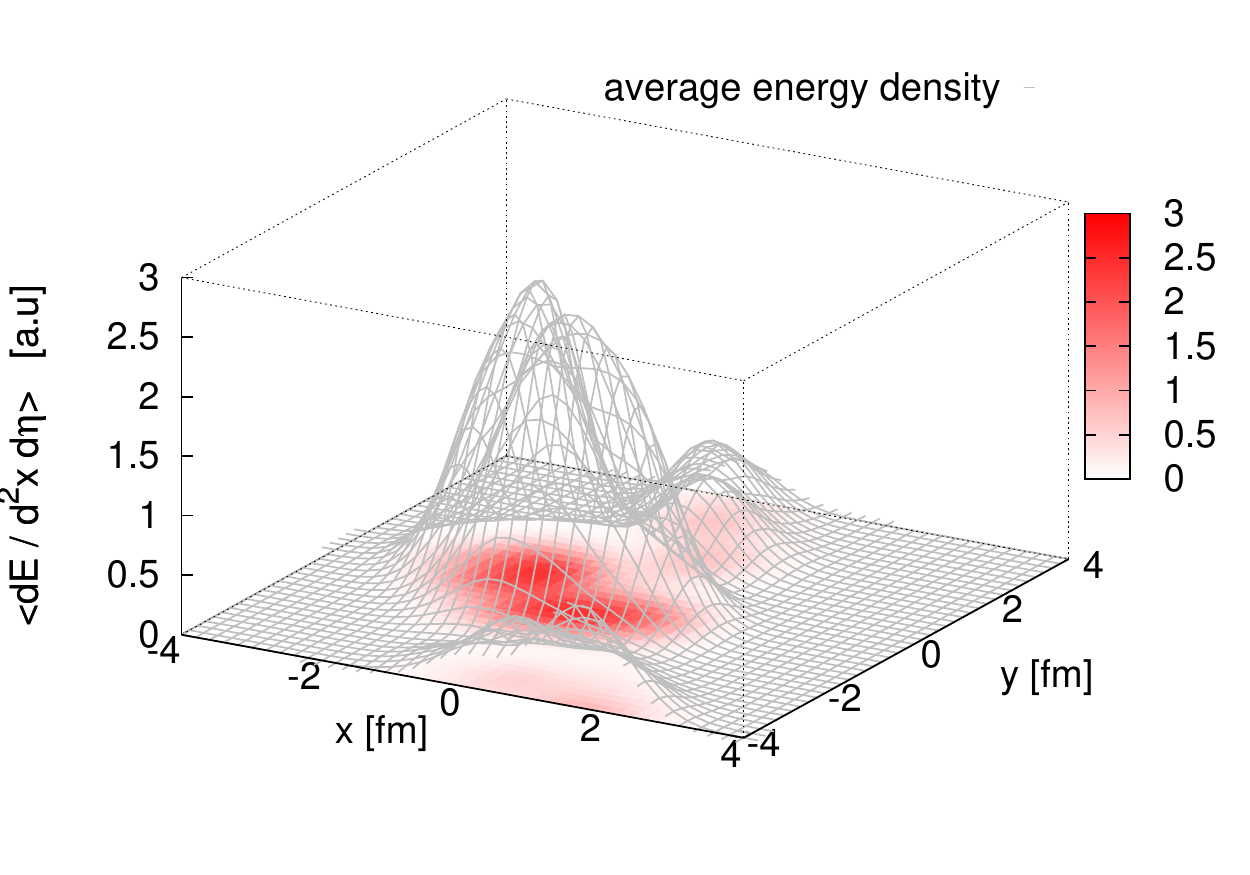}
\includegraphics[width=0.6\textwidth]{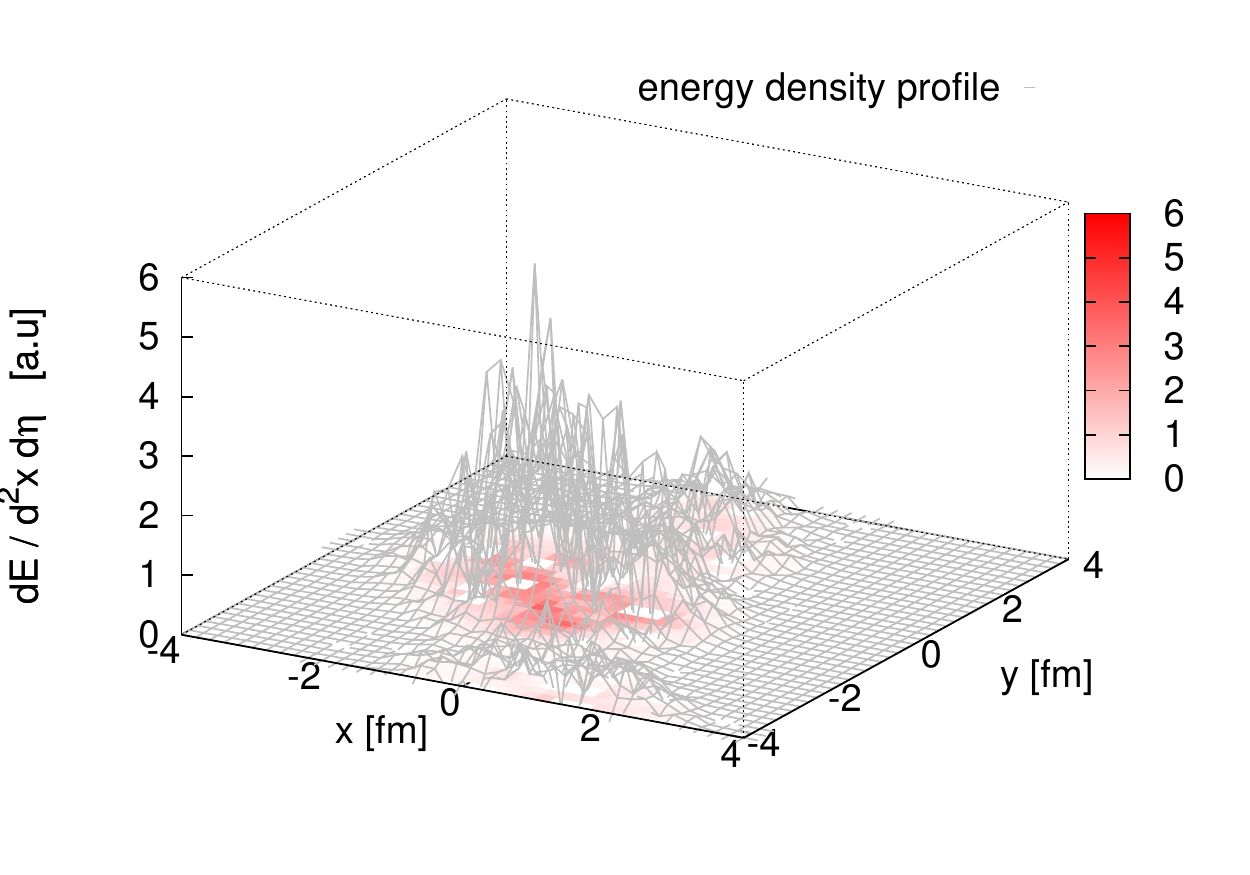}
\includegraphics[width=0.6\textwidth]{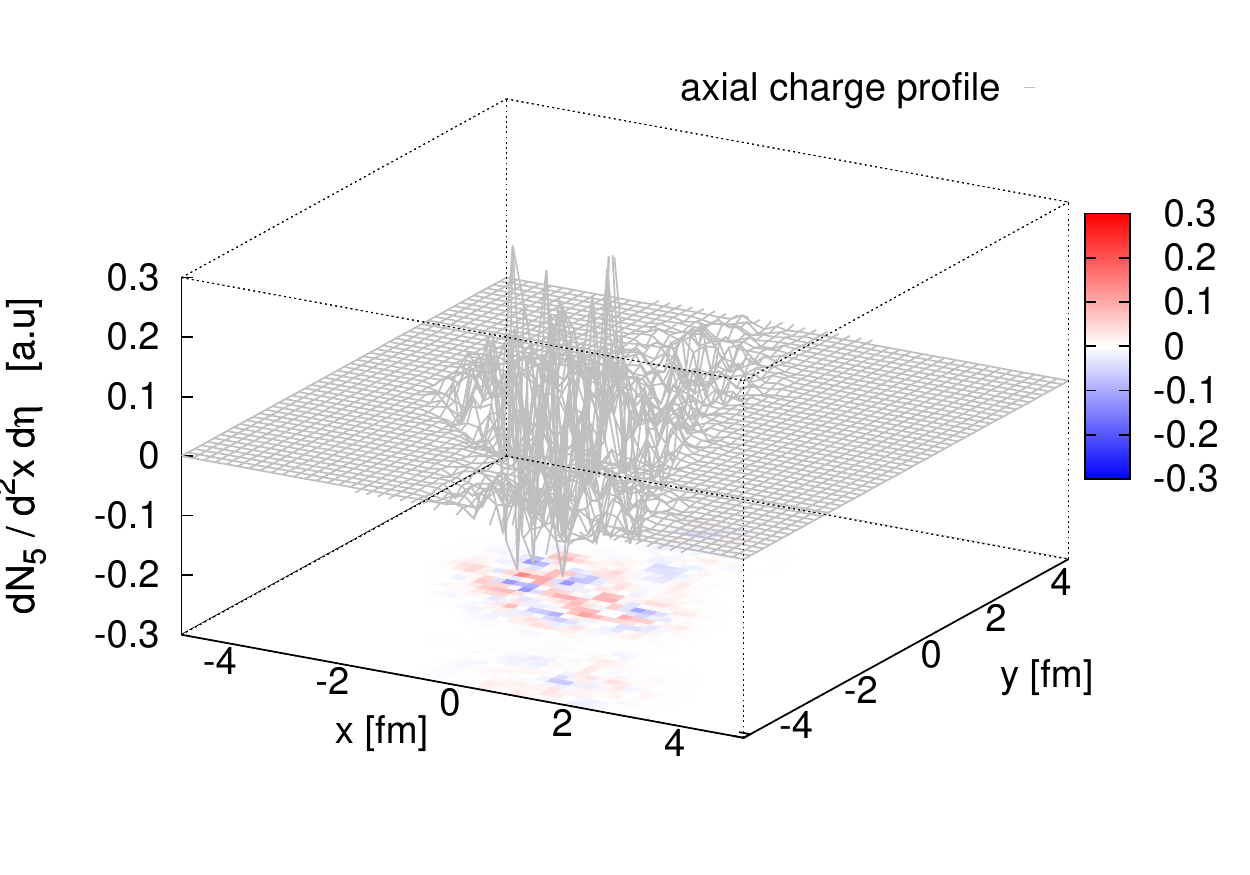}
\caption{\label{fig:ChernSimonsMap} Energy density and axial charge profiles for a peripheral Pb$+$Pb event ($b=11.4~$fm | $N_\text{part}=56$) -- the typical correlation length of axial charge distribution is on the order of the inverse saturation scale chosen as $\qs^2=2~$GeV$^2$.}
\end{center}
\end{figure}

We also note that our result in Eq.~(\ref{eq:KeyResult}) can directly be used to model the initial conditions of the axial charge density $\frac{\ud N_5}{\ud^2\xt \ud\eta}$ in anomalous hydrodynamics or other calculations that attempt to relate anomalous transport phenomena to experimental measurements. 
We start from the assumption that on an event-by-event basis one knows the average energy density profile $\langle \epsilon(\xt) \rangle$ as a function of the transverse coordinates, e.g. from a Monte Carlo Glauber model. 
This energy density profile should be thought of as the ``average'' energy density in the the sense that color charge fluctuations at the scale $\qs$ are not included. The fluctuations at longer length scales, such as those resulting from fluctuations of the positions of nucleons inside the nucleus, should be averaged over separately as an external input to our calculation. 
Assuming that the energy density profile is sampled at a discrete set of points $\xt$ in the transverse plane, one straightforward way to generate configurations of the axial charge distribution with a given two-point correlation function
\begin{eqnarray}
\left\langle \frac{\ud N_5}{\ud^2\xt \ud\eta} \frac{\ud N_5}{\ud^2\yt \ud\eta} \right \rangle=C(\xt,\yt)~\langle \epsilon(\xt)\rangle~\langle \epsilon(\yt)\rangle
\end{eqnarray}
as in Eq.~(\ref{eq:KeyResult}) is to perform a Cholesky decomposition of the correlation function
\begin{eqnarray}
C(\xt,\yt)=\sum_{\zt} L(\xt,\zt) L^{T}(\zt,\yt)\;.
\end{eqnarray}
By sampling individual configurations of the axial charge distribution according to 
\begin{eqnarray}
\frac{\ud N_5}{\ud^2\xt \ud\eta}=\langle \epsilon(\xt)\rangle \sum_{\zt} \xi(\zt)~ L(\xt,\zt)
\end{eqnarray}
where $\xi(\zt)$ are uncorrelated random numbers with zero mean $\langle \xi(\zt) \rangle=0$ and unit variance $\langle \xi(\zt) \xi(\zt') \rangle=\delta_{\zt,\zt'}$, it is then straightforward to verify that the correlation function is correctly reproduced on average. Similarly, our result in Eq.~(\ref{eq:EnergyFluctuationResult}) can also be used to include additional sub-nucleonic $\qs$-scale fluctuations of the energy density $\langle \epsilon(\xt) \epsilon(\yt) \rangle - \langle \epsilon(\xt) \rangle \langle \epsilon(\yt) \rangle$ on top of the average energy density profile $\langle \epsilon(\xt) \rangle$ by following the same procedure outlined above. This provides a simplistic way to include the kind of $\qs$-scale energy density fluctuations that are present in the IPglasma model~\cite{Schenke:2012wb,Schenke:2012hg}, although the analytic expressions used here are just approximations of the full numerical result. We emphasize that the procedure can be applied to any model or parametrization for the initial energy density at very early times $\tau \ll 1/\qs$. Even if the initial average energy density does not come from an explicit saturation model calculation, one can estimate the corresponding saturation scale by solving for $\qs$ from the initial energy density $\left< \varepsilon\right> \approx \frac{1}{g^2} \frac{\nc^2}{\cf} \qs^4$. 

We illustrate this procedure in Fig.~\ref{fig:ChernSimonsMap}, with the example of a peripheral Pb+Pb event. Based on the average energy density profile obtained from the $\textsc{T}_\textsc{R}\textsc{ENTO}$ event generator~\cite{Moreland:2014oya} shown in the first panel of Fig.~\ref{fig:ChernSimonsMap}, we include fluctuations of the energy density and axial charge distribution following the above procedure. Despite the fact that average energy density profile is rather smooth, with typical variations on size scales $\sim$~fm, sub-nucleonic fluctuations give rise to fluctuations of the energy density at length scales $\sim 1/\qs$ as can be seen from the central panel of Fig.~\ref{fig:ChernSimonsMap}. Similarly, variations of the axial charge distribution due to Glasma flux tubes occur on microscopic length scales with a characteristic size $\sim 1/\qs$. However, due to the approximate boost invariant nature of the Glasma fields, these structures are elongated in rapidity. It will be interesting to see from phenomenological calculations whether such small structures can have a sizeable effect on hadronic observables. In order to facilitate the use of our result in this context, we provide the source code for generating axial charge distributions as in Fig.~\ref{fig:ChernSimonsMap}  as supplementary material.

\section{Conclusions \& Perspectives}
\label{sec:conc}
Based on known analytic solutions for the Glasma fields we calculated energy and axial charge fluctuations at early times $\tau\lesssim 1/\qs$ after the collision of heavy nuclei at high energies.  Our calculation generalizes the earlier work of~\cite{Muller:2011bb} to a more general structure for the gluon distribution and, more importantly, to derive an expression for the Chern-Simons correlator.  Generally, we find that the expressions for energy and axial charge fluctuations in Eqs.~\nr{eq:corr} involve the correlation function of two Weisz\"{a}cker-Williams (WW) gluon distributions, represented as a correlator of eight light-like Wilson lines for each nucleus. We evaluated this correlation function in the ``Glasma graph'' approximation, where the relevant double parton distribution is factorized into a product of single parton distributions. Based on previous calculations \cite{Lappi:2015vta}, we expect the Glasma graph approximation to be quite close to the full result. Extending this calculation to the full nonlinear Gaussian treatment would require working out an eight-point function of Wilson lines  in the similar way as the four-point function in  Appendix~\ref{app:quadrupole}. Based on the complexity of the expressions it appears unlikely that this could be done analytically, but a numerical evaluation similar to the recent one in \cite{Dusling:2017dqg} should certainly be feasible. We also note that, based on our primary interest of applications to the collision of large nuclei, we neglected some more subtle effects related to position-momentum correlations in the gluon distribution (see e.g. \cite{Hagiwara:2016kam,Hagiwara:2017ofm}), which may be interesting to investigate in further applications to small systems.

Our result in Eqs.~\nr{eq:epsepscorr} and \nr{eq:nunucorr} expresses fluctuations of the energy density and axial charge in terms of the Bessel moments of the unpolarized ($G^{(1)}$) and linearly polarized ($h^{(1)}_{\bot}$) transverse momentum dependent gluon distributions. Interestingly, we find that the spin structure enters in a different way in the final expressions. In particular the two-point correlator of the Chern-Simons term is sensitive to the difference of Bessel moments of unpolarized and linearly polarized distributions. Evaluating the Weisz\"{a}cker-Williams distributions in a simple saturation model (GBW), we also provided explicit estimates in Eq.~(\ref{eq:EnergyFluctuationResult}) and (\ref{eq:KeyResult}) for energy density and axial charge fluctuations at early times. 

In view of possible phenomenological applications of our result, we provided a practical algorithm to use our result to implement quantitatively the axial-charge density fluctuations in the Glasma. Of course, this relation relies on a rough treatment of the time dependence of the Chern-Simons charge in the Glasma (c.f. Sec.~\ref{sec:fluxtubes}) and ultimately a full classical Yang-Mills calculation including dynamical fermions along the lines of \cite{Gelis:2005pb,Gelis:2015eua,Mace:2016shq} as well as a realistic geometry will be needed. Even with this approximation our result should, however, enable a better control of the initial conditions for anomalous hydrodynamics simulations or other calculations that are needed to relate these ideas to experimental measurements. We caution, however, that axial charge changing processes e.g. due to sphaleron transitions or thermal fluctuations of the field strength tensor continue to take place throughout the entire space-time evolution of the Quark-Gluon Plasma. Clearly such effects should also be included in realistic model calculations and further theoretical progress will be required. 

We finally note that several calculations similar to ours have been performed for momentum space gluon correlations based on the Glasma graph approximation~\cite{Dumitru:2008wn,Gelis:2009wh,Dusling:2009ni,Dumitru:2010iy,Dusling:2012iga,Dusling:2012wy,Dusling:2013qoz}. However, these calculations are performed in the dilute limit and do not give access to the linear polarization structure of the gluon distribution. Focusing only on the coordinate space correlation structures of the fields at $\tau \lesssim 1/\qs$ enables us to do a calculation in  a manifestly gauge invariant way and more cleanly elucidate the role of the gluon polarization. The relation between our present work and the Glasma graph calculations of ridge correlations is explained in more detail in Sec.~\ref{sec:momspace}.

\begin{acknowledgments}
We thank A.~Dumitru, A.~Kovner, M.~Lublinsky and R.~Venugopalan for insightful discussions on subjects closely related to this work, J.-F. Paquet for providing the $\textsc{T}_\textsc{R}\textsc{ENTO}$ event and B.~M{\"u}ller and A.~Sch{\"a}fer for correspondence concerning Ref.~\cite{Muller:2011bb}. We thank the CERN-TH department for hospitality during the time when this work has initiated. Support by the U.S. Department of Energy, Office of Science, Nuclear Physics under Grant. No. DE-FG02-97ER41014 (S.S.) as well as by the Academy of Finland, projects 267321 and 303756 (T.L.) and by the European Research Council, Grant ERC-2015-CoG-681707 (T.L.) is gratefully acknowledged by the authors.
\end{acknowledgments}

\appendix
\section{Evaluation of Weisz\"{a}cker-Williams distribution in Gaussian models}
\label{app:EvaluationGaussianModel}
We start by decomposing the gluon fields $\alpha^{i}_{\xt}$ over the Lie Algebra
\begin{eqnarray}
\alpha^{i}_{\xt}=\frac{i}{g} U_{\xt} \partial^{i} U_{\xt}^{\dagger}\;, \qquad \alpha^{i}_{\xt}=\alpha^{i,a}_{\xt} t^{a}\;, \qquad \alpha^{i,a}_{\xt}= \frac{2i}{g} \tr\Big( t^{a}~U_{\xt} \partial^{i} U_{\xt}^{\dagger} \Big) 
\end{eqnarray}
such that the Weizs\"{a}cker-Williams distribution is given by
\begin{eqnarray}
\frac{g^2 (\nc^2-1)}{2} W^{ik}_{(U)}(\xt,\yt)= 2~\left\langle \tr\Big( t^{a}~U_{\xt} i\partial^{i} U_{\xt}^{\dagger} \Big) \tr\Big( t^{a}~U_{\yt} i\partial^{k} U_{\yt}^{\dagger} \Big) \right\rangle
\end{eqnarray}
By re-expressing the derivatives in terms of new coordinates $\xd,\yd$ and making use of  the $SU(\nc)$ Fierz identity
\begin{eqnarray}
t^{a}_{ij} t^{a}_{kl}=\frac{1}{2} \delta_{il} \delta_{jk}-\frac{1}{2\nc} \delta_{ij} \delta_{kl}
\end{eqnarray}
the relevant correlation function of Wilson lines then take the form
\begin{eqnarray}
\frac{g^2 (\nc^2-1)}{2} W^{ik}_{(U)}(\xt,\yt)=  i\partial^{i}_{\xd}~i\partial^{k}_{\yd}~ \left. \left\langle~\tr\Big( U_{\xt} U_{\xd}^{\dagger} U_{\yt} U_{\yd}^{\dagger} \Big) - \frac{1}{\nc}  \tr\Big( U_{\xt} U_{\xd}^{\dagger}\Big) \tr\Big(U_{\yt} U_{\yd}^{\dagger} \Big) \right\rangle \right|_{\xd=\xt,~\yd=\yt}\;. \nonumber \\
\end{eqnarray}
Clearly the second term vanishes upon taking the derivative and setting coordinates $\xd=\xt$ and $\yd=\yt$ equal to each, as $U_{\xt} \partial^{i} U_{\xt}^{\dagger}$ are elements of the Lie algebra and thus traceless. We are then left with evaluating the first term involving the quadrupole correlator.\\

\subsection{Evaluation of the Wilson line correlators in Gaussian model}
We perform the Gaussian averaging of the correlators of  Wilson line, by expressing the usual Gaussian integral over color charges in terms of a stochastic process in the evolution variable $z \in [0,1]$ such that the Wilson lines at $z=0$ are given by $V_{\xt}(z=0)=1$ and in each step
\begin{eqnarray}
\partial_{z} V_{\xt} = V_{\xt} \Big(+igt^{a} \xi^{a}_{\xt}(z)\Big)\;,
\end{eqnarray} 
where $\xi^{a}_{\xt}$ are stochastic variables with
\begin{eqnarray}
\left<\xi^{a}_{\xt}(z)\xi^{b}_{\yt}(z')\right>=\frac{1}{g^2\cf} \delta^{ab} \lambda_{\xt\yt}\delta(z-z')\;,
\end{eqnarray}
where $C_{F}=(\nc^2-1)/(2\nc)$ is the fundamental Casimir. Starting for simplicity with the dipole operator, we can then evaluate 
\begin{eqnarray}
\partial_{z} \left\langle \frac{1}{\nc} \tr\Big( U_{\xt} U_{\xd}^{\dagger} \Big)\right\rangle=G_{\xt\xd}~\left\langle \tr\Big( U_{\xt} U_{\xd}^{\dagger} \Big)\right\rangle
\end{eqnarray}
where we introduced the correlation functions
\begin{eqnarray}
G_{\xt\xd}=\lambda_{\xt\xd} - \frac{1}{2}\lambda_{\xt\xt} -\frac{1}{2}\lambda_{\xd\xd}\;,
\end{eqnarray}
such that the dipole correlator is simply given by
\begin{eqnarray}
 D_{\xt\xd}=\frac{1}{\nc}\left\langle \tr\Big( U_{\xt} U_{\xd}^{\dagger} \Big)\right\rangle= \exp\Big(G_{\xt\xd}\Big)\;.
\end{eqnarray}

\subsection{Quadrupole \& Dipole-Dipole correlators}
\label{app:quadrupole}
Similarly for the quadrupole, we obtain upon use of the $SU(\nc)$ Fierz identity the evolution equation
\begin{eqnarray}
\partial_{z} \tr\Big( U_{\xt} U_{\xd}^{\dagger} U_{\yt} U_{\yd}^{\dagger} \Big)  &=& \Big(G_{\xt\xd} +G_{\yt\yd} - \frac{1}{\nc^2-1} T_{\xt\xd,\yt\yd} \Big)  \tr\Big( U_{\xt} U_{\xd}^{\dagger} U_{\yt} U_{\yd}^{\dagger} \Big)  \\
&& + \frac{1}{2\cf}  T_{\xt\xd,\yt\yd}  \tr\Big( U_{\xt} U_{\yd}^{\dagger} \Big) \tr\Big( U_{\xd}^{\dagger} U_{\yt}  \Big)\;, \nonumber 
\end{eqnarray}
where the transition function $T_{\xt\xd,\yt\yd}$ given by
\begin{eqnarray}
T_{\xt\xd,\yt\yd}=\lambda_{\xt\yd}+\lambda_{\yt\xd}-\lambda_{\xt\yt}-\lambda_{\xd\yd}=G_{\xt\yd}+G_{\yt\xd}-G_{\xt\yt}-G_{\xd\yd}
\end{eqnarray}
This has to be supplemented with the evolution equation for the dipole-dipole correlator
\begin{eqnarray}
\partial_{z}  \tr\Big( U_{\xt} U_{\yd}^{\dagger} \Big) \tr\Big( U_{\xd}^{\dagger} U_{\yt}  \Big) &=& \Big(G_{\xt\yd} +G_{\yt\xd} - \frac{1}{\nc^2-1} T_{\xt\yd,\yt\xd} \Big) \tr\Big( U_{\xt} U_{\yd}^{\dagger} \Big) \tr\Big( U_{\xd}^{\dagger} U_{\yt}  \Big)  \\
&&+ \frac{1}{2\cf}  T_{\xt\yd,\yt\xd} \tr\Big( U_{\xt} U_{\xd}^{\dagger} U_{\yt} U_{\yd}^{\dagger} \Big)\;, \nonumber 
\end{eqnarray}
with the transition function $T_{\xt\yd,\yt\xd}$ given by
\begin{eqnarray}
T_{\xt\yd,\yt\xd}=G_{\xt\xd}+G_{\yt\yd}-G_{\xt\yt}-G_{\xd\yd} \; .
\end{eqnarray}
We obtain the coupled set of evolution equations
\begin{eqnarray}
\partial_{z} \begin{pmatrix}  \tr\Big( U_{\xt} U_{\yd}^{\dagger} \Big) \tr\Big( U_{\xd}^{\dagger} U_{\yt}  \Big)  \\ \tr\Big( U_{\xt} U_{\xd}^{\dagger} U_{\yt} U_{\yd}^{\dagger} \Big) \end{pmatrix}=
M(\xt,\xd,\yt,\yd) \begin{pmatrix}  \tr\Big( U_{\xt} U_{\yd}^{\dagger} \Big) \tr\Big( U_{\xd}^{\dagger} U_{\yt}  \Big)  \\ \tr\Big( U_{\xt} U_{\xd}^{\dagger} U_{\yt} U_{\yd}^{\dagger} \Big) \end{pmatrix} \nonumber \\
\end{eqnarray}
where the evolution operator $M(\xt,\xd,\yt,\yd)$ takes the form
\begin{eqnarray}
M(\xt,\xd,\yt,\yd) =\begin{pmatrix} 	G_{\xt\yd} +G_{\yt\xd} - \frac{1}{\nc^2-1} T_{\xt\yd,\yt\xd}  & \frac{1}{2\cf}  T_{\xt\yd,\yt\xd} \\ \frac{1}{2\cf}  T_{\xt\xd,\yt\yd}  & G_{\xt\xd} +G_{\yt\yd} - \frac{1}{\nc^2-1} T_{\xt\xd,\yt\yd} \end{pmatrix}\;.
\end{eqnarray}
Hence, the relevant correlation function can be obtained as
\begin{eqnarray}
 \tr\Big( U_{\xt} U_{\xd}^{\dagger} U_{\yt} U_{\yd}^{\dagger} \Big) = \begin{pmatrix} 0  \\ 1 \end{pmatrix}^{T} \exp\Big( M(\xt,\xd,\yt,\yd) \Big) \begin{pmatrix} \nc^2 \\ \nc \end{pmatrix}\;.
\end{eqnarray}
Of course, for this simple example we could easily calculate the full expression as done in \cite{Dominguez:2011wm}. However, for our purpose it is more useful to first take the derivatives and set the coordinates $\xd=\xt$ and $\yd=\yt$ equal to each other, such that the relevant expression
\begin{eqnarray}
\frac{g^2 (\nc^2-1)}{2} W^{ik}_{\alpha}(\xt,\yt)&=&i\partial_{\xd}^{i} i\partial_{\yd}^{k}~\left.\tr\Big( U_{\xt} U_{\xd}^{\dagger} U_{\yt} U_{\yd}^{\dagger} \Big) \right|_{\xd=\xt,~\yd=\yt}
\end{eqnarray}
greatly reduces in complexity to
\begin{equation}
\frac{g^2 (\nc^2-1)}{2} W^{ik}_{\alpha}(\xt,\yt)=\begin{pmatrix} 0 \\ -1 \end{pmatrix}^{T} \partial_{\xd}^{i} \partial_{\yd}^{k}  \left. \exp\Big( M(\xt,\xd,\yt,\yd) \Big) \right|_{\xd=\xt,~\yd=\yt} \begin{pmatrix} \nc^2 \\ \nc \end{pmatrix} 
\; .
\end{equation} 
Specifically, denoting the evolution matrix  and its derivatives as
\begin{eqnarray}
M_{\xt\yt}=M(\xt,\xt,\yt,\yt)\;, \qquad 
M^{(i,0)}_{\xt\yt}=\left. \partial_{\xd}^{i} M(\xt,\xd,\yt,\yd) \right|_{\xd=\xt,~\yd=\yt}\;, \\
M^{(0,k)}_{\xt\yt}=\left. \partial_{\yd}^{k} M(\xt,\xd,\yt,\yd) \right|_{\xd=\xt,~\yd=\yt}\;,
M^{(i,k)}_{\xt\yt}=\left.\partial_{\xd}^{i} \partial_{\yd}^{k} M(\xt,\xd,\yt,\yd)\right|_{\xd=\xt,~\yd=\yt} \nonumber 
\end{eqnarray}
 the derivative of the matrix exponential is given by
\begin{multline}
\label{eq:Correlator}
\partial_{\xd}^{i} \partial_{\yd}^{k}  \left. \exp\Big( M(\xt,\xd,\yt,\yd) \Big) \right|_{\xd=\xt,~\yd=\yt} = \int_{0}^{1} \ud s \exp\Big(s~M_{\xt\yt}\Big)~M^{(i,k)}_{\xt,\yt}~\exp\Big((1-s)~M_{\xt\yt}\Big) 
\\
\quad +\int_{0}^{1} \ud s \int_{0}^{1} \ud t \exp\Big(st~M_{\xt\yt}\Big)~s M^{(i,0)}_{\xt,\yt}~\exp\Big((1-t)s~M_{\xt\yt}\Big)~M^{(0,k)}_{\xt,\yt}~\exp\Big((1-s)~M_{\xt\yt}\Big)  \\
\quad +\int_{0}^{1} \ud s \int_{0}^{1} \ud t \exp\Big(s~M_{\xt\yt}\Big)~M^{(0,k)}_{\xt,\yt}~\exp\Big((1-s)t~M_{\xt\yt}\Big)
\\  \quad \times
(1-s) M^{(i,0)}_{\xt,\yt}~\exp\Big((1-s)(1-t)~M_{\xt\yt}\Big)\;. 
\end{multline}
Evaluating the matrix elements according to
\begin{eqnarray}
G_{\xt\yt}=G_{\yt\xt}\;, \qquad G_{\xt\xt}=G_{\yt\yt}=0\;, \qquad T_{\xt\xt,\yt\yt}=0\;, \qquad T_{\xt\yt,\yt\xt}=-2G_{\xt\yt}
\end{eqnarray}
we obtain
\begin{eqnarray}
M_{\xt\yt}=\begin{pmatrix} 	\frac{2 \nc^2}{\nc^2-1} G_{\xt\yt}  & -\frac{1}{\cf}  G_{\xt\yt} \\ 0  & 0 \end{pmatrix}\;.
\end{eqnarray}
such that
\begin{eqnarray}
\exp\Big( q~M_{\xt\yt} \Big) =  \begin{pmatrix} e^{q\frac{2 \nc^2}{\nc^2-1} G_{\xt\yt}}  & \frac{1}{\nc} \Big(1- e^{q\frac{2 \nc^2}{\nc^2-1} G_{\xt\yt}}  \Big)  \\ 0  & 1 \end{pmatrix}\;.
\end{eqnarray}
Similarly, using the relations
\begin{eqnarray}
\partial_{\xd}^{i} \left. G_{\xt\xd}\right|_{\xd=\xt}=0\;, \qquad \partial_{\xd}^{i} \left. G_{\xd\yt}\right|_{\xd=\xt}=\partial_{\xd}^{i}  \left. G_{\yt\xd}\right|_{\xd=\xt}=G^{(i)}_{\xt\yt} \\
\partial_{\xd}^{i} \left.T_{\xt\xd,\yt\yd} \right|_{\xd=\xt}=G^{(i)}_{\yt\xt}-G^{(i)}_{\yd\xt}\;, \qquad  \partial_{\xd}^{i} \left.T_{\xt\yd,\yt\xd} \right|_{\xd=\xt}=-G^{(i)}_{\xt\yd}
\end{eqnarray}
we obtain the derivative of the evolution operator as
\begin{eqnarray}
\partial_{\xd}^{i} \left.M(\xt,\xd,\yt,\yd)\right|_{\xd=\xt} =\begin{pmatrix} 	G^{(i)}_{\yt\xt} + \frac{1}{\nc^2-1} G^{(i)}_{\xt\yd}  & -\frac{1}{2\cf}  G^{(i)}_{\xt\yd}\\ \frac{1}{2\cf}  \Big(G^{(i)}_{\yt\xt}-G^{(i)}_{\yd\xt} \Big)  &  -\frac{1}{\nc^2-1} \Big(G^{(i)}_{\yt\xt}-G^{(i)}_{\yd\xt} \Big) \end{pmatrix}\;.
\end{eqnarray}
such that the relevant expressions are given by
\begin{eqnarray}
M^{(i,0)}_{\xt\yt} &=& \begin{pmatrix} 	\frac{\nc^2}{\nc^2-1}G^{(i,0)}_{\xt\yt}  & -\frac{1}{2\cf}  G^{(i,0)}_{\xt\yt}\\ 
0  & 0 \end{pmatrix}\;.\nonumber \\
M^{(0,k)}_{\xt\yt} &=& \begin{pmatrix} 	\frac{\nc^2}{\nc^2-1}G^{(0,k)}_{\xt\yt}  & -\frac{1}{2\cf}  G^{(0,k)}_{\xt\yt}\\ 
0  & 0 \end{pmatrix}\;.\nonumber \\
M^{(i,k)}_{\xt\yt}&=&\begin{pmatrix} 	+ \frac{1}{\nc^2-1} G^{(i,k)}_{\xt\yt}  & -\frac{1}{2\cf}  G^{(i,k)}_{\xt\yt}\\ -\frac{1}{2\cf}  G^{(i,k)}_{\xt\yt}   & +\frac{1}{\nc^2-1} G^{(i,k)}_{\xt\yt} \end{pmatrix}\;. 
\end{eqnarray}
Based on the explicit form of $M^{(i,0)}_{\xt\yt}$ and $M^{(0,k)}_{\xt\yt}$ with vanishing entries in the second line, the first derivative terms in Eq.~(\ref{eq:Correlator}) vanish upon the projection onto the final state and hence do not contribute to the quadrupole operator. Collecting everything and performing the integrals we finally obtain 
\begin{eqnarray}
\frac{g^2 (\nc^2-1)}{2} W^{ik}_{(U)}(\xt,\yt)=\cf \frac{G^{(i,k)}_{\xt\yt}}{G_{\xt\yt}} \Big( e^{\frac{2 \nc^2}{\nc^2-1} G_{\xt\yt}} -1 \Big)\;.
\end{eqnarray}

\bibliography{spires}
\bibliographystyle{JHEP-2modlong}

\end{document}